\documentclass[preprint,review,12pt]{elsarticle}
\usepackage{amssymb}
\usepackage{amsmath} 
\usepackage[ruled]{algorithm2e}
\usepackage{colortbl,booktabs} 
\usepackage{tabularx}
\usepackage{threeparttable}
\usepackage{multicol,multirow}
\usepackage{graphicx}
\usepackage{subfigure} 
\usepackage{enumerate}
\usepackage{float}
\usepackage{graphicx}
\usepackage{caption}
\usepackage [font=footnotesize,labelfont=bf]{caption}
\usepackage{rotating}
\usepackage{tikz}
\usepackage[colorlinks,linkcolor=blue]{hyperref}
\usepackage{color}
\definecolor{R}{rgb}{1, 0.0, 0.0}
\definecolor{G}{rgb}{0.0, 1, 0.0}
\definecolor{B}{rgb}{0.0, 0.0, 1}

\begin{document}
\begin{frontmatter}
	\title{Towards high-order consistency and convergence 
		   of conservative SPH approximations}
	\author{Bo Zhang}
	\ead{bo.zhang.aer@tum.de}
	\author{Nikolaus Adams}
	\ead{nikolaus.adams@tum.de}
	\author{Xiangyu Hu\corref{mycorrespondingauthor}}
	\cortext[mycorrespondingauthor]{Corresponding author.}
	\ead{xiangyu.hu@tum.de}
	\address{TUM School of Engineering and Design,
		Technical University of Munich,\\
		85748 Garching, Germany}
	\begin{abstract}
		Smoothed particle hydrodynamics (SPH) offers distinct 
		advantages for modeling many engineering problems, yet 
		achieving high-order consistency in its conservative 
		formulation remains to be addressed.
		While zero- and higher-order consistencies can be 
		obtained using particle-pair differences and the kernel 
		gradient correction (KGC) approaches, respectively, for 
		SPH gradient approximations, their applicability for 
		discretizing conservation laws in practical simulations 
		is limited due to their non-conservative feature.
		Although the standard anti-symmetric SPH approximation 
		is able to achieve conservative zero-order consistency 
		through particle relaxation, its straightforward 
		extensions with the KGC fail to satisfy either zero- 
		or higher-order consistency.
		In this paper, we propose the reverse KGC (RKGC) formulation, 
		which is conservative and able to precisely satisfy both 
		zero- and first-order consistencies when particles are 
		relaxed based on the KGC matrix.
		Extensive numerical examples show that the new formulation 
		considerably improves the accuracy of the Lagrangian SPH 
		method. 
		In particular, it is able to resolve the long-standing 
		high-dissipation issue for simulating free-surface flows.
		Furthermore, with fully relaxed particles, it enhances the 
		accuracy of the Eulerian SPH method even when the ratio 
		between the smoothing length and the particle spacing is 
		considerably reduced.
		Indeed, the reverse KGC formulation holds the potential 
		for the extension to even higher-order consistencies. 
		However, addressing the corresponding particle relaxation 
		problem remains a pending challenge.
	\end{abstract}
\end{frontmatter}
%
%
\section{Introduction}\label{introduction}
As a mesh-free method, smoothed particle hydrodynamics 
(SPH), initially proposed by Lucy \cite{lucy1977numerical} 
and Gingold \& Monaghan \cite{gingold1977smoothednew} for 
astrophysical applications, has demonstrated significant 
success across a wide range of scientific problems.  
These include fluid dynamics \cite{monaghan1994simulating, 
	hu2006multi, zhang2017weakly}, 
solid dynamics \cite{benz1995simulations, 
	rabczuk2003simulation, peng2019meshfree},
and fluid-structure interaction \cite{ye2019smoothed, 
	gotoh2018state, zhang2021multi}, among others.
The SPH approximation operates on the principle of 
reconstructing the continuous field and its spatial 
derivatives from a collection of discrete particles,
each possessing individual properties, through a Gaussian-like 
smoothing kernel function with compact support 
\cite{fulk1994numerical, fulk1996analysis}.
It generally encounters two different types of errors 
that amalgamate to the overall truncation error 
\cite{quinlan2006truncation, litvinov2015towards}.
The first is the smoothing error determined by the 
kernel function where the leading moments vanish.
This error arises due to the discrepancy between the 
smoothing approximations and the exact values. 
The second is the integration error, characterized 
by the non-vanishing of leading moments due to the 
particle approximation.
For typical SPH kernel functions, such as the cubic B-spline 
\cite{monaghan1992smoothed} and the Wendland kernel 
\cite{maz'ia2007approximate} are second-order accuracy, 
given that only the first moment vanishes, corresponding 
to first-order consistency.
With the given kernel function, Quinlan et al. 
\cite{quinlan2006truncation} observed that the overall 
truncation error generally decreases or exhibits 
consistency with increased resolution only when 
$h/\Delta{x}$, i.e., the ratio between the smoothing 
length $h$ and the particle spacing $\Delta{x}$, is 
large, suggesting sufficient small integration error.     
However, achieving this condition leads to an excessive 
number of particles within the kernel compact support 
and results in an extremely high computational cost.

Therefore, different approaches have been proposed 
to minimize integration error or improve consistency  
with a computationally acceptable value of $h/\Delta{x}$,
typically less than 1.5 \cite{oger2007improved}.
Besides that zero-order consistency can be easily achieved 
by using particle-pair differences for gradient approximations, 
high-order consistencies can be attained through kernel 
gradient correction (KGC) \cite{randles1996smoothed} and 
various similar approaches, such as corrective smoothed 
particle method (CSPM) \cite{chen1999corrective}, 
reproducing kernel particle method (RKPM) 
\cite{liu1995reproducing, liu1995reproducing2},
finite particle method \cite{liu2006restoring}, 
modified SPH method (MSPH) \cite{batra2004analysis, 
	zhang2004modified, sibilla2015algorithm, nasar2021high}, 
moving least squared (MLS) \cite{atluri1999analysis}
and many others \cite{flyer2011radial, king2020high, trask2017high}.    
While these approaches are able to achieve consistencies 
for the SPH approximation of the gradient and/or Laplacian 
operators, their application to the discretization of 
physical conservation laws still faces a significant 
challenge of non-conservation. 
Specifically, conservative discretization necessitates 
an anti-symmetric form between particle pairs, a condition 
that these methods could not appropriately satisfy.

The first approach to achieve both zero-order 
consistency and conservation involves implementing 
the particle relaxation based on constant background 
pressure \cite{litvinov2015towards, zhu2021cad} 
before applying anti-symmetric SPH approximations.
Since the particle relaxation is computationally 
expensive for the Lagrangian SPH method, the alternative 
one-step consistency correction, such as the 
transport-velocity formulation \cite{adami2013transport, 
	zhang2017generalized}, based on the same principle, 
has also been proposed to enhance the consistency. 
Although conservation properties always hold a high priority 
\cite{oger2007improved}, there is still an expectation 
for high-order consistency in SPH so that it can be 
utilized as a generally effective numerical method.
To address this, approaches have been developed to enhance 
numerical accuracy and consistency by integrating KGC 
with anti-symmetric formulations \cite{nasar2019eulerian, 
	bonet1999variational, de2022new} and/or particle 
relaxation \cite{ren2023efficient}.
These methods include utilizing the average correction 
matrix \cite{vila2005sph, zago2021overcoming, liang2023study} 
and implementing separate corrections for each particle pair 
\cite{oger2007improved, huang2022development}, etc.
For some problems, while these conservative KGC formulations 
have demonstrated improved results compared to standard SPH 
methods without KGC, as will be shown later, they not only 
fail to achieve high-order consistency but also lose 
zero-order consistency, even with particle relaxation.

In this study, we introduce the reverse KGC (RKGC) 
formulation designed for conservative SPH approximations. 
In contrast to the prior conservative KGC formulations, 
RKGC is able to precisely fulfill both zero- and 
first-order consistencies of the gradient operator.
The formulation incorporates a particle relaxation driven 
by the KGC matrix and is distinguished into two parts: 
the first part addresses zero-order consistency and 
vanishes during particle relaxation, while the second 
part ensures first-order consistency and accurately 
reproduces the linear gradient. 
The formulation notably improves numerical accuracy in 
Lagrangian SPH simulations. 
In particular, it exhibits very good energy conservation 
properties and resolves the long-standing high-dissipation 
issue for the SPH simulation of free-surface flows.
Moreover, since these consistencies can be strictly 
imposed in the Eulerian SPH method, even when employing 
a reduced smoothing length, the formulation still has 
the potential to yield results with improved accuracy.

In the following sections, Section \ref{preliminary} presents 
the approximation of gradients in the SPH method and its 
application in discretizing governing equations.
Section \ref{RKGC} introduces the RKGC formulation, detailing 
the particle relaxation and the transport-velocity formulation 
based on the KGC matrix. 
Section \ref{convergenceanalysis} conducts corresponding error 
and convergence analyses.
Following this, Section \ref{numericalexamples} presents 
extensive numerical examples that highlight the benefits gained 
from the proposed method.
In Section \ref{extension}, we extend the RKGC formulation to 
second-order consistency, with a yet-to-be-addressed condition 
on the particle relaxation.
Finally, Section \ref{conclusion} summarizes the key findings 
and outlines of future research.  
%
%
\section{Preliminary}\label{preliminary}
\subsection{Gradient approximation}
\label{gradientapproximation}
In SPH, the kernel approximation for the gradient of 
a smooth field $\psi\left({\boldsymbol{\rm r}}\right)$
can be expressed through a two-stage approach
\begin{equation}
	\nabla\psi\left(\boldsymbol{\rm r}\right)\approx
	\int_{\Omega}\nabla\psi\left({\boldsymbol{\rm r}^{*}}
	\right)W\left(\boldsymbol{\rm r}-\boldsymbol{\rm r}^{*},
	h\right)d\boldsymbol{\rm r}^{*}=
	-\int_{\Omega}\psi\left({\boldsymbol{\rm r}^{*}}
	\right)\nabla W\left(\boldsymbol{\rm r}-
	\boldsymbol{\rm r}^{*},h\right)
	d\boldsymbol{\rm r}^{*},
	\label{SPHapproximation}
\end{equation}
where $W\left(\boldsymbol{\rm r}, h\right)$ is the kernel 
function scaled by the smoothing length $h$.
While the first stage introduces smoothing errors by the 
kernel function, the second stage entails the integration 
by parts, assuming the kernel function vanishes at the 
boundary of compact support.
Through Taylor expansion, for Eq. \eqref{SPHapproximation}, one 
can easily find that the zero-order consistency condition is
\begin{equation}
	\int_{\Omega}\nabla W\left(\boldsymbol{\rm r}-
	\boldsymbol{\rm r}^{*},h\right)d\boldsymbol{\rm r}^{*}=0,
	\label{kernelproperty}
\end{equation}
and the first-order consistency condition is
\begin{equation}
	-\int_{\Omega}\left(\boldsymbol{\rm r}^{*}-
	\boldsymbol{\rm r}\right)\otimes\nabla W
	\left(\boldsymbol{\rm r}-\boldsymbol{\rm r}^{*},
	h\right)d\boldsymbol{\rm r}^{*}=\mathbf{I},
	\label{firstorderconsistencyintegration}
\end{equation} 
where $\mathbf{I}$ represents the identity matrix.
Note that, with zero-order consistency condition, one can 
rewrite the kernel approximation in two equivalent forms:
\begin{multline}
	\nabla\psi\left(\boldsymbol{\rm r}\right)=
	\int_{\Omega}\left(\psi\left({\boldsymbol{\rm r}}
	\right) - \psi\left({\boldsymbol{\rm r}^{*}}\right)\right)
	\nabla W\left(\boldsymbol{\rm r}-\boldsymbol{\rm r}^{*},
	h\right)d\boldsymbol{\rm r}^{*} \\
	\equiv 
	-\int_{\Omega}\left(\psi\left({\boldsymbol{\rm r}}
	\right) + \psi\left({\boldsymbol{\rm r}^{*}}\right)\right)
	\nabla W\left(\boldsymbol{\rm r}-\boldsymbol{\rm r}^{*},
	h\right)d\boldsymbol{\rm r}^{*}.
	\label{SPHgradientapproximations}
\end{multline}
By introducing particle summation, the first approximation 
in Eq. \eqref{SPHgradientapproximations} can be further 
approximated at an SPH particle $i$ as
\begin{equation}
	\nabla\psi_{i} = \sum_{j}
	\psi_{ij} \nabla W_{ij}V_{j},
	\label{strong}
\end{equation}
where $V_{j}$ is the volume of the neighbor particles within 
the support, and the particle-pair difference $\psi_{ij}=
\psi_{i}-\psi_{j}$ is adopted.
This form is often referred to as a symmetric or non-conservative form.
Similarly, the second approximation in Eq. 
\eqref{SPHgradientapproximations} can be further approximated as
\begin{equation}
	\nabla\psi_{i} = -\sum_{j}
	\left(\psi_{i}+\psi_{j}\right) \nabla W_{ij}V_{j},
	\label{weak}
\end{equation}
where the particle-pair sum is employed.
This form, known as the anti-symmetric or conservative form, 
ensures conservation properties and is commonly chosen in classic 
SPH methods for the discretization of physical conservation laws.

For the non-conservative form of Eq. \eqref{strong}, zero-order 
consistency is automatically satisfied as the particle-pair 
difference is used.
To achieve first-order consistency, one requires that the 
approximation of Eq. \eqref{firstorderconsistencyintegration} 
satisfying
\begin{equation}
	-\sum_{j}\boldsymbol{\rm r}_{ij}\otimes
	\nabla W_{ij}V_{j} = \mathbf{I}.
	\label{firstorderconsistency}
\end{equation}
To precisely fulfill the above condition, the KGC approach 
\cite{randles1996smoothed}, introducing a correction matrix 
$\mathbf{B}_{i}$ to adjust the gradient of the kernel 
function, can be employed, so that one has 
\begin{equation}
	-\sum_{j}\boldsymbol{\rm r}_{ij}\otimes
	\mathbf{B}_{i}\nabla W_{ij}V_{j}=\mathbf{I}, 
	\quad \mathbf{B}_{i}=\left(-\sum_{j}
	\mathbf{r}_{ij}\otimes\nabla  
	W_{ij}V_{j}\right)^{-1}.
	\label{correctionmatrix}
\end{equation}
With the KGC, Eq. \eqref{strong} is modified into 
\begin{equation}
	\nabla\psi_{i} = \sum_{j}
	\psi_{ij} \mathbf{B}_{i}\nabla W_{ij}V_{j}.
	\label{strong-corrected}
\end{equation}
Note that, introducing $\mathbf{B}_{i}$ does not affect
the zero-order consistency of Eq. \eqref{strong-corrected}.
Also note that, although the non-conservation form is not 
desirable for the discretization of physical conservation laws, 
Eq. \eqref{strong-corrected} is often used when the conservation 
is not a primary concern because it can reproduce the linear 
gradient and achieve second-order accuracy.

In the conservative form of Eq. \eqref{weak}, where the 
particle-pair sum other than the difference is used, 
the zero-order consistency condition becomes nontrivial as  
\begin{equation}
	\sum_{j}\nabla W_{ij}V_{j}=0.
	\label{zeroorderconsistency}
\end{equation}
Litvinov et al. \cite{litvinov2015towards} proposed a particle 
relaxation process driven by a constant background pressure 
assuming invariant particle volume.
After the particles are settled or fully relaxed, Eq. 
\eqref{zeroorderconsistency} is satisfied for the zero-order 
consistency.
To achieve first-order constancy, as a straightforward 
extension for Eq. \eqref{weak}, one may expect that 
\begin{equation}
	\nabla\psi_{i}=-\sum_{j}
	\left(\psi_{i}\mathbf{B'}_{i}+\psi_{j}\mathbf{B'}_{j}\right)
	\nabla W_{ij}V_{j},
	\label{expected-kgc}
\end{equation}
where $\mathbf{B'}_{i}$ and $\mathbf{B'}_{j}$ are some correction 
matrices for particles $i$ and $j$, is able to reproduce linear 
gradient similar to Eq. \eqref{strong-corrected}.
Since how to obtain these correction matrices is not straightforward,
various attempts based on the origin KGC matrix for non-conservative 
form have been carried out.
One widely spread formulation, introduced by Oger et al. 
\cite{oger2007improved}, is expressed as,
\begin{equation}
	\nabla\psi_{i}=-\sum_{j}
	\left(\psi_{i}\mathbf{B}_{i}+\psi_{j}\mathbf{B}_{j}\right)
	\nabla W_{ij}V_{j},
	\label{skgc}
\end{equation}
where the KGC matrix is applied for each of the particle pair. 
\subsection{Weakly compressible SPH (WCSPH)}
The governing equations in the Lagrangian framework for viscous 
flows consist of the mass and momentum conservation equations, 
written as
\begin{equation}
	\begin{cases}
		\vspace{5pt}
		\displaystyle\dfrac{\text{d}\rho}{\text{d}t} 
		=-\rho\nabla\cdot\boldsymbol{\rm v}\\
		\displaystyle\dfrac{\text{d}\boldsymbol
		{\rm v}}{\text{d}t} 
		=-\dfrac{1}{\rho}\nabla p+\nu\nabla^{2}
		\boldsymbol{\rm v}+\boldsymbol{\rm g}
	\end{cases},
	\label{wcsphgoverningequation}
\end{equation}
where $\rho$ represents density, $\boldsymbol{\rm v}$ velocity, 
$p$ pressure, $\nu$ kinematic viscosity, $\boldsymbol{\rm g}$ 
gravity and $\text{d}(\bullet)/\text{d}t=\partial(\bullet)
/\partial t+\boldsymbol{\rm v}\cdot\nabla(\bullet)$ refers 
to the material derivative.
An artificial equation of state (EOS) for weakly compressible 
flows is used to close Eq. \eqref{wcsphgoverningequation} as 
\begin{equation}
	p= {c_0}^2\left(\rho-\rho_{0}\right).
	\label{wcspheos}
\end{equation}
Here, $\rho_{0}$ is the initial density, and $c_{0}$ denotes 
the artificial sound speed. 
Setting $c_{0} = 10 U_{max}$, where $U_{max}$ represents the 
anticipated maximum fluid speed, fulfills the weakly compressible 
assumption where the density variation remains around $1\%$.

The Riemann-SPH method \cite{zhang2017weakly} is employed here 
to discretize Eq. \eqref{wcsphgoverningequation}.
Subsequently, the continuity and momentum equations are 
approximated as
\begin{equation}
	\begin{cases}
		\vspace{5pt}
		\displaystyle\dfrac{\text{d}\rho_{i}}{\text{d}t}
		=2\rho_{i}\sum_{j}\left(\boldsymbol{\rm v}_i-
		\boldsymbol{\rm v^*}\right)\cdot \nabla W_{ij}V_{j}\\
		\displaystyle\dfrac{\text{d}\boldsymbol{\rm v_{i}}}
		{\text{d}t}= -\dfrac{2}{m_{i}}\sum_{j}P^{*}\nabla 
		W_{ij}V_{i}V_{j}+2\sum_{j}\dfrac{\nu}{\rho_{i}}
		\frac{\boldsymbol{\rm v}_{ij}}{r_{ij}} \dfrac{\partial 
			W}{\partial{r_{ij}}}V_{j}+\boldsymbol{\rm g}_{i}
	\end{cases},
	\label{wcsphformulation}
\end{equation}
where $\nabla W_{ij}= \frac{\partial W_{ij}}{\partial r_{ij}}
\boldsymbol{\rm e}_{ij}$, and $\boldsymbol{\rm e}_{ij}=\frac{
	\boldsymbol{\rm r}_{ij}}{{\rm r}_{ij}}$ is the unit vector. 
The particle-pair velocity $\boldsymbol{\rm v}^{*}$ and pressure
$P^{*}$, respectively, are solutions obtained from the Riemann 
problem constructed along the interacting line of each pair of 
particles.
Note that the particle-pair pressure $P^{*}$ leads to 
an anti-symmetric form and hence momentum conservation.
With a linearised Riemann solver, the solutions can be 
computed as
\begin{equation}
	\begin{cases}
		\vspace{5pt}
		\boldsymbol{\rm v}^{*}=\overline{\boldsymbol{
		\rm v}}_{ij}+\left(U^{*}-\overline{U}_{ij}\right)
		\boldsymbol{\rm e}_{ij},\quad
		U^{*}=\overline{U}_{ij} +\dfrac{1}{2} 
		\dfrac{p_{ij}}{\rho_{0}c_{0}}\\
		P^{*}=\overline{p}_{ij}+\dfrac{1}{2}\beta
		\rho_{0}c_{0}U_{ij}
	\end{cases}
\label{linearsolution}.
\end{equation}
Here, $\overline{(\bullet)}_{ij}=\left[(\bullet)_i+(\bullet)_j
\right]/2$ represents particle-pair average, $\overline{U}_{ij}=
-\overline{\boldsymbol{\rm v}}_{ij}\cdot {\rm e}_{ij}$ and 
$U_{ij} = -\boldsymbol{\rm v}_{ij}\cdot {\rm e}_{ij}$, represent the 
particle-pair average and difference of the particle velocity along 
the interaction line, respectively, and the low-dissipation limiter 
is defined as $\beta=\min\left(3\max\left(U_{ij}/c_{0},0\right),1\right)$.
Additionally, it should be noted that the particle-pair pressure 
$P^*$ in Eq. \eqref{linearsolution} comprises two main components: 
a non-dissipative term denoted by $\overline{p}_{ij}$, and a 
dissipative term derived from the differences between particle pairs.
\subsection{Eulerian SPH (ESPH)}
The conservation equations for weakly compressible flows in 
the Eulerian framework are expressed as
\begin{equation}
	\dfrac{\partial\boldsymbol{\rm U}}{\partial t}+ 
	\nabla\cdot\boldsymbol{\rm F}
	\left(\boldsymbol{\rm U}\right)=0,
	\label{esphgoverningequation}
\end{equation}
where $\boldsymbol{\rm U}=(\rho, \rho\boldsymbol{\rm v})$ denotes 
the vector of conserved variables, and $\boldsymbol{\rm F}\left
(\boldsymbol{\rm U}\right)$ represents the corresponding fluxes.
Following the methodology outlined in Refs. \cite{vila1999particle, 
	wang2023extended}, the Eulerian SPH discretization of Eq. 
\eqref{esphgoverningequation} can be expressed in an anti-symmetric 
or conservative form as
\begin{equation}
	\begin{cases}
		\vspace{5pt}
		\displaystyle\dfrac{\partial}{\partial t}\left(
		\rho_{i}V_{i}\right)+2\sum_{j}
		\left(\rho\boldsymbol{\rm v}\right)^{*}_{E,ij}
		\cdot\nabla W_{ij} V_{i}V_{j} =0\\
		\displaystyle\dfrac{\partial}
		{\partial t}\left(\rho_{i}\boldsymbol{\rm v}_{i}V_{i}
		\right)+2\sum_{j}\left[\left(\rho
		\boldsymbol{\rm v}\otimes\boldsymbol{\rm v}\right)+
		p\mathbf{I}\right]^{*}_{E,ij}\cdot\nabla W_{ij}V_{i}V_{j}=0
	\end{cases}.
\end{equation}
Here, terms $\left(\bullet\right)^{*}_{E,ij}$ denote numerical 
fluxes for each particle pair, determined by solutions of the 
Riemann problem \cite{vila1999particle}.
The HLL Riemann solver \cite{harten1983upstream, 
	rieper2010dissipation} incorporating a low-dissipation limiter 
\cite{zhang2017weakly} is adopted here to solve the Riemann problem.
The solutions as numerical fluxes can be written as
\begin{equation}
	\boldsymbol{\rm F}^{*}=\frac{1}{2}
	\overline{\boldsymbol{\rm F}}_{ij}+
	\beta\left(\dfrac{1}{2}
	\dfrac{S_{R}+S_{L}}{S_{R}-S_{L}}
	\boldsymbol{\rm U}_{ij}+
	\dfrac{S_{R}S_{L}}{S_{R}-S_{L}}
	\boldsymbol{\rm F}_{ij}\right).
	\label{hllsolution}
\end{equation}
Here, $S_{L}$ and $S_{R}$ represent the wave speeds estimated in 
the left and right regions of the Riemann problem (see Ref. 
\cite{harten1983upstream} for more details), respectively, 
satisfying the assumption of $S_L\le 0\le S_R$ and $S_l\ne S_R$, 
which is validated for weakly compressible flows.
The same low-dissipation limiter $\beta$ as in Eq. 
\eqref{linearsolution} is utilized to handle the dissipative 
component of the numerical fluxes.
Again, the non-dissipative component in Eq. \eqref{hllsolution} 
is given by the particle-pair average of the physical fluxes.
%
%
\section{Reverse KGC formulation}\label{RKGC}
\subsection{Consistent correction method}\label{derivation}
Although it has been shown that the straightforward employment of
the KGC matrix, as outlined in Eq. \eqref{skgc}, is able to obtain 
improved results for some problems \cite{nasar2019eulerian, 
	huang2022development, ren2023efficient} compared to the 
standard conservative formulation in Eq. \eqref{weak}, it loses 
both zero- and first-order consistencies. 
The formulation of Eq. \eqref{skgc} can be rewritten as
\begin{equation}
	\nabla\psi_{i}=-\psi_{i}\sum_{j}\left(\mathbf{B}_{i}+
	\mathbf{B}_{j}\right)\nabla W_{ij}V_{j}+
	\sum_{j}\psi_{ij} \mathbf{B}_{j}\nabla W_{ij}V_{j}
	\label{nofirstorderconsistency}.
\end{equation}
It can be found that the first term on the right-hand side (RHS), 
mimicking Eq. \eqref{zeroorderconsistency}, gives the zero-order 
consistency condition when incorporating the KGC, and the second 
term, again mimicking Eq. \eqref{strong-corrected}, attempts to 
reproduce the linear gradient.
However, the first term generally does not vanish even after the 
particle relaxation driven by constant pressure 
\cite{litvinov2015towards} due to the modification by the KGC matrix.
In addition, the second term is different from the original form 
as the KGC matrix of neighboring particles is employed and, 
consequently, does not guarantee first-order consistency either.
The same issues also arise in other corrected formulations 
\cite{vila2005sph, zago2021overcoming, liang2023study}.

Therefore, we modify Eq. \eqref{nofirstorderconsistency} by using 
the KGC matrix of particle $i$ as
\begin{equation}
	\nabla\psi_{i}=-\psi_{i}\sum_{j}\left(\mathbf{B}_{i}+
	\mathbf{B}_{j}\right)\nabla W_{ij}V_{j}+\sum_{j}\psi_{ij}
	\mathbf{B}_{i}\nabla W_{ij}V_{j}
	\label{rkgcexplain},
\end{equation}
so that the second term is the same as Eq. \eqref{strong-corrected} 
and achieves first-order consistency.
If the first term also vanishes for achieving zero-order consistency, 
the entire formulation satisfies both consistencies at the same time.
Similar to employing constant pressure for particle relaxation, here, 
we can consider $\mathbf{B}_{i}$ and $\mathbf{B}_{j}$ as some geometric 
stresses dependent on the particle distribution and use them to drive 
the particle relaxation.
It is easy to find that after the particle is settled or fully 
relaxed under such KGC particle relaxation, the first term vanishes.
Note that, Eq. \eqref{rkgcexplain} can be simplified into the 
following anti-symmetric form
\begin{equation}
	\nabla\psi_{i}=-\sum_{j}\left(\psi_{i}\mathbf{B}_{j}+
	\psi_{j}\mathbf{B}_{i}\right)\nabla_{i}W_{ij}V_{j}.
	\label{rkgc}
\end{equation}
Comparing Eq. \eqref{rkgc} with Eq. \eqref{skgc}, one can find that 
the only difference is that, in the new formulation satisfying both 
zero- and first-order consistencies, the KGC matrix is employed 
reversely with respect to particle pair $i$ and $j$.
Therefore, Eq. \eqref{rkgc} is denoted as a reverse KGC (RKGC) 
formulation.

In this work, the RKGC formulation is employed in the discretization 
of the momentum conservation equation for the Lagrangian SPH method, 
and both the mass and momentum conservation equations for the Eulerian 
SPH method by replacing the original particle-pair average for the 
non-dissipative terms in the Riemann solutions as the form of 
$\overline{(\bullet\mathbf{B})}_{ij}=\left[(\bullet)_i\mathbf{B}_{j}
+(\bullet)_j \mathbf{B}_{i}\right]/2$ in Eqs. \eqref{linearsolution} 
and \eqref{hllsolution}.
\subsection{KGC particle relaxation and transport-velocity formulation}
\label{relaxation}
The original relaxation method (denoted as P relaxation or PR), as 
detailed in Ref. \cite{litvinov2015towards, zhu2021cad}, operates 
by executing particle relaxation driven by a constant background 
pressure to obtain the zero-order consistency condition as outlined 
in Eq. \eqref{zeroorderconsistency}. 
It iteratively adjusts particle positions to rectify zero-order 
integration errors, with the correction in each step determined by
\begin{equation}
	\Delta\mathbf{r}_{i}=\alpha\left(\Delta{x}\right)^2
	\sum_j\nabla W_{ij}V_{j}
	\label{Prelaxation}.
\end{equation}
Here, the parameter $\alpha = 0.2$ is chosen to ensure numerical 
stability until the relaxation error reaches a sufficiently small 
value.

To incorporate the proposed RKGC formulation, we introduce the 
KGC relaxation (denoted as B relaxation or BR), where the particle 
relaxation is driven by the geometric stress or the KGC matrix, to 
obtain the zero-order consistency as the first term in Eq. 
\eqref{rkgcexplain} suggested.
Similar to P relaxation, the iterative correction on particle 
positions in each step is accordingly modified into
\begin{equation}
	\Delta\mathbf{r}_i = \alpha\left(\Delta{x}\right)^2\sum_j
	(\mathbf{B}_{i}+\mathbf{B}_{j})\nabla W_{ij}V_j.
	\label{Brelaxation}
\end{equation}
Note that the KGC matrix for each particle is recomputed by 
Eq. \eqref{correctionmatrix} before each iteration step.
Similar to the P relaxation, this B relaxation could also provide 
a uniformly distributed particle distribution, with body-fitted 
particles for complex geometries.

In the Lagrangian SPH method, the original transport-velocity 
formulation \cite{adami2013transport}, where a single correction step 
according to Eq. \eqref{Prelaxation} is used (for example in each 
advection time step of the dual-time step SPH method \cite{zhang2020dual}) 
to enhance (without pursing full) zero-order consistency and avoid 
particle clustering when negative pressure presents.
Similarly, a KGC transport-velocity formulation can be proposed in the 
same fashion, except the single correction step is replaced with that of 
Eq. \eqref{Brelaxation}.
Note that both transport-velocity formulations only slightly modify the 
particle positions without modifying the velocity or the total momentum 
of the entire system.  
%
%
%
\section{Error and convergence analysis}
\label{convergenceanalysis}
The accuracy and convergence of the SPH gradient operators in 
conservative form without correction (NKGC) in Eq. \eqref{weak},
with the original straightforward KGC (SKGC) in Eq. \eqref{skgc}, 
and with the reverse KGC (RKGC) in Eq. \eqref{rkgc}, are investigated.
A circle domain with the radius $R=1.0$ is considered, and a 
scalar field is initialized within the domain by the function  
\begin{equation}
	\psi\left(x\right)= e^{-10x^2}.
\end{equation} 
As in Ref. \cite{quinlan2006truncation}, the error is measured with the 
non-dimensionalized $L_{2}$ norm defined as
\begin{equation}
	\epsilon L_{2}=\sqrt{\dfrac{1}{N_{t}}\left(\sum_{i}|
	\nabla\psi^{ANA}_{i}-\nabla \psi^{SPH}_{i}|^2\right)}
	\label{l2normerror},
\end{equation}
where $\psi^{ANA}_{i}$ and $\psi^{SPH}_{i}$ represent the analytical 
and numerical solutions, and $N_{t}$ is the total number of particles 
within the domain of interest (sufficiently away from the boundary).  
The C2 Wendland kernel is utilized to conduct tests on 
lattice-distributed (without body-fit particles at the boundary) and 
relaxed particle distributions with both P and B relaxations.
The convergence criterion for relaxation is set at a maximum 
zero-order consistency residue of $1\times 10^{-5}$. 
The particle spacing $\Delta{x}$ ranges from $0.2$ to $0.0125$, while 
the smoothing lengths $h$ of $1.3\Delta x$, $1.15\Delta x$, and 
$0.8\Delta x$ are used to study the convergence with decreasing $h/\Delta x$.

\begin{figure}[htbp]
	\centering
	\vspace{-2.5cm}
	\subfigure[]{\includegraphics[width=.75\textwidth]
		{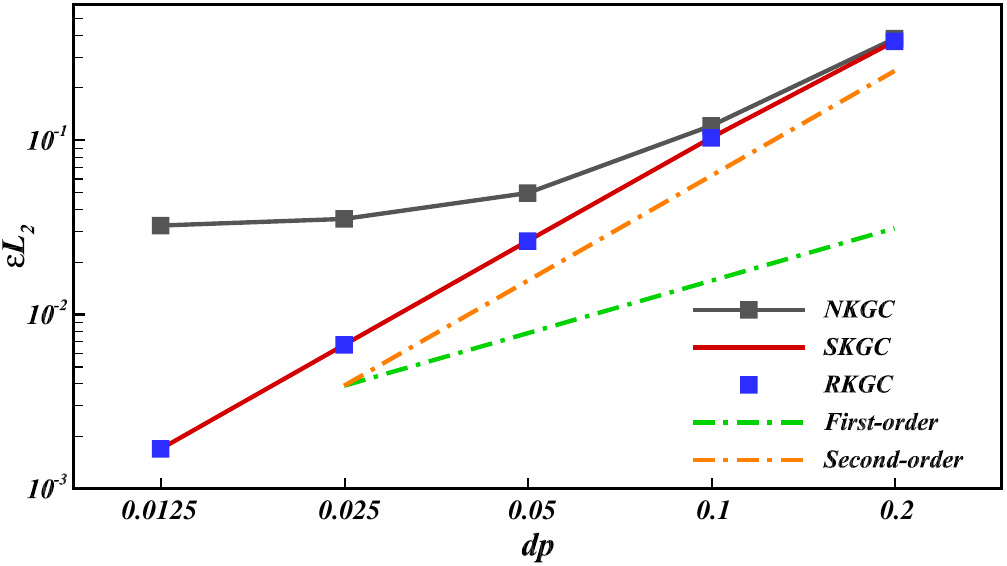}\label{latticedistribution}}
	\vspace{-0.3cm}
	\subfigure[]{\includegraphics[width=.75\textwidth]
		{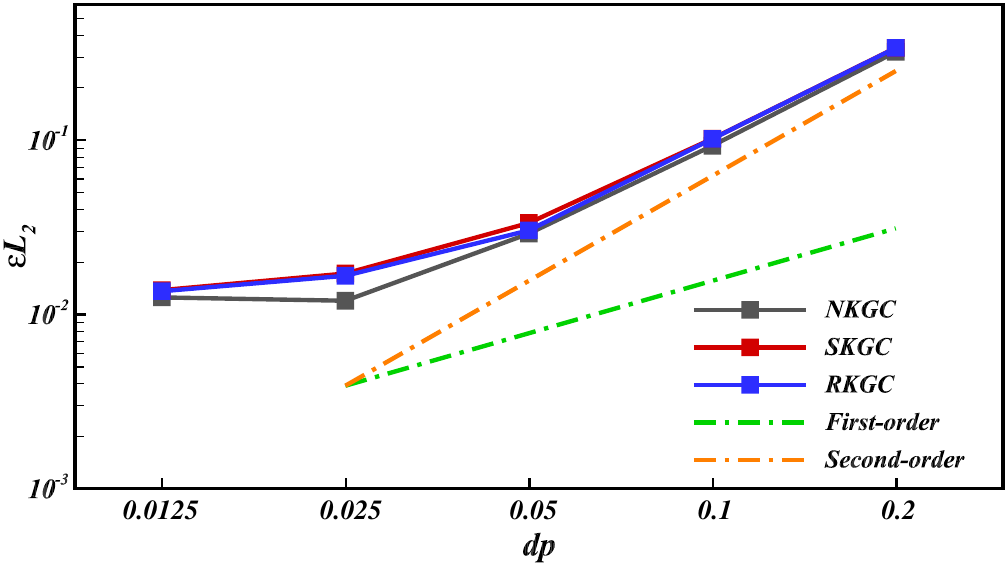}\label{pressurerelaxation}}
	\vspace{-0.3cm}
	\subfigure[]{\includegraphics[width=.75\textwidth]
		{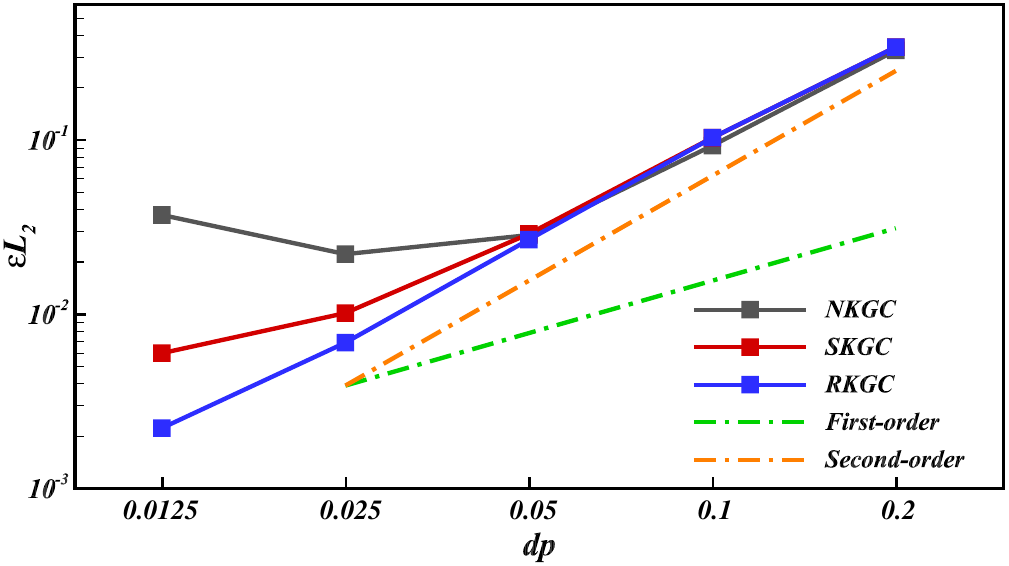}\label{correctionmatrixrelaxation}}
	\caption{Convergence study of conservative approximations 
		of the gradient at the $h=1.3\Delta x$. (a) Lattice 
		distribution; (b) P relaxation; (c) B relaxation.}
	\label{gradientapproximationconvergence}
\end{figure}

Fig. \ref{gradientapproximationconvergence} presents the convergences 
with increasing resolutions at $h=1.3\Delta x$. 
For lattice-distributed particles, as illustrated in 
Fig. \ref{latticedistribution}, both corrected formulations achieve 
second-order convergence compared to the typical second-order-to-saturation 
behavior of NKGC. 
For the particles after the P relaxation, as depicted in Fig. 
\ref{pressurerelaxation}, all formulations exhibit the second-order-to-saturation 
behavior, as the integration errors dominant at high-resolution independent 
of the kernel corrections.
For particles after the B relaxation, as shown in Fig. 
\ref{correctionmatrixrelaxation} only RKGC maintains second-order convergence as 
expected from the analysis in section \ref{derivation}. 
At high resolutions, SKGC degrades to first-order as it fails to reproduce 
the linear gradient, and NKGC even suffers from increased error.   

\begin{figure}[htbp]
	\centering
	\subfigure[]{\includegraphics[width=.75\textwidth]
		{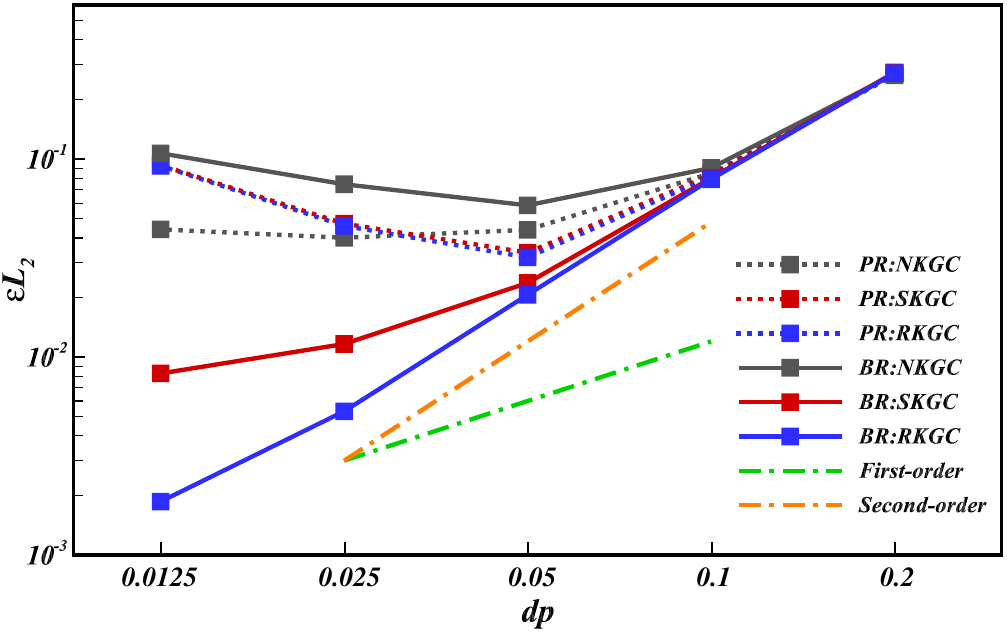}\label{smoothinglength1.15}}
	\subfigure[]{\includegraphics[width=.75\textwidth]
		{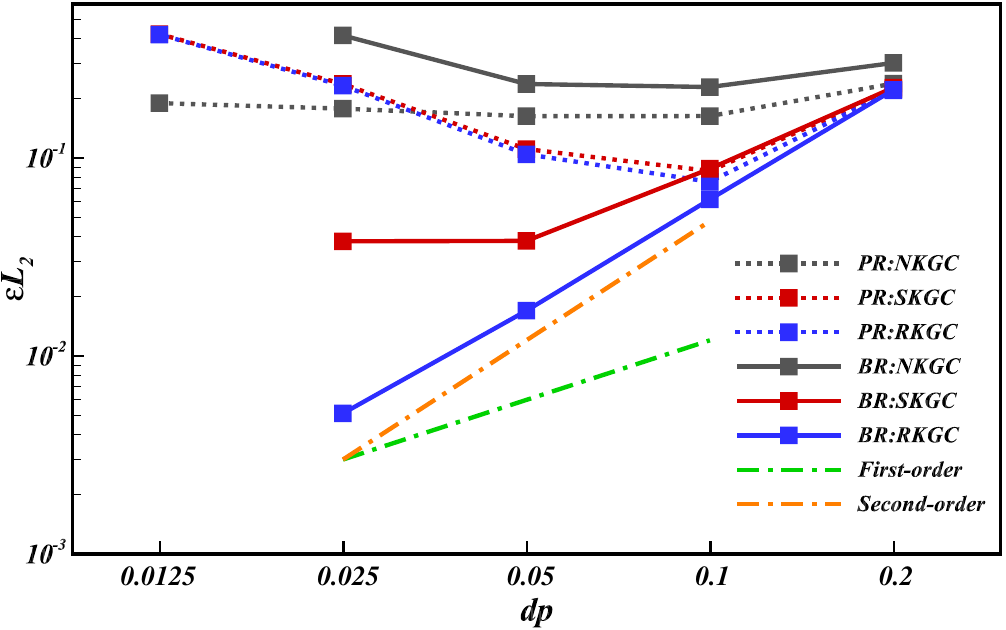}\label{smoothinglength0.8}}
	\caption{Convergence study of conservative approximations 
		of the gradient at the reduced $h/\Delta x$ values. 
		(a) $h=1.15 \Delta x$; (b) $h=0.8 \Delta x$.}
	\label{convergestudywithvarysmoothinglength}
\end{figure}

Furthermore, as depicted in Fig. \ref{convergestudywithvarysmoothinglength}, 
it is observed that, for RKGC, reducing the $h/\Delta x$ does not affect 
the convergence rate as long as the B relaxation is applied. 
This is not out of expectation since Eq. \eqref{rkgcexplain} is not 
explicitly dependent on smoothing length.
In contrast, SKGC and NKGC suffer from serious degeneration or even 
increased error at high resolutions, no matter whether P or B relaxation 
is applied.
Note that the data at $\Delta x = 0.0125$ for $h=0.8\Delta x$ is missing 
for B relaxation as shown in Fig. \ref{smoothinglength0.8}.
This is because, under the present straightforward relaxation stepping 
as given in Section \ref{relaxation}, the B relaxation is not able to 
reach the convergence criterion. Such difficulty actually indicates the 
implicit limitation of the RKGC formulation.
However, such limitation is generic and more demanding for high-order 
consistencies as shown latter, and even not unexpected since a basic 
requirement of consistent particle approximation is sufficient overlapping 
between the kernel supports of neighboring particles \cite{koumoutsakos2005multiscale}. 
%
%
%
\section{Numerical examples}\label{numericalexamples}
In this section, the proposed RKGC formulation is applied to the 
WCSPH and ESPH methods with fully relaxed particles for the latter. 
Again, the C2 Wendland kernel is utilized, with the smoothing length 
set to be $h=1.3\Delta x$ if not specifically stated.
\subsection{Taylor-Green vortex flow at Re=100}
The incompressible Navier-Stokes equation offers an analytical time-dependent
solution for this periodic array of vortices in a unit square domain as
\begin{equation}
	\begin{cases}
		u\left(x,y,t\right)=-Ue^{bt}\cos\left(2\pi x\right)
		\sin\left(2\pi y\right)\\
		v\left(x,y,t\right)=Ue^{bt}\sin\left(2\pi x\right)
		\cos\left(2\pi y\right)
	\end{cases}.
\end{equation}
This solution serves as the initial velocity distribution at 
$t=0$ and acts as a benchmark to assess the simulation accuracy.
The decay rate of the velocity field is determined by 
$b=-8\pi^{2}/Re$, where $Re=\rho UL/\eta$ represents the Reynolds 
number derived from the fluid density $\rho$, the maximum initial 
velocity $U$, the domain size $L$, and the viscosity $\eta$. 
In the current simulations, a domain size of $L=1$ is employed, 
with periodic boundary conditions applied in both coordinate 
directions.
The maximum initial flow speed is set at $U=1$, and the Reynolds 
number is $Re=100$.
\subsubsection{WCSPH results}
\begin{figure}[htb!]
	\makebox[\textwidth][c]{
		\subfigure[]{\includegraphics[width=.5\textwidth]
			{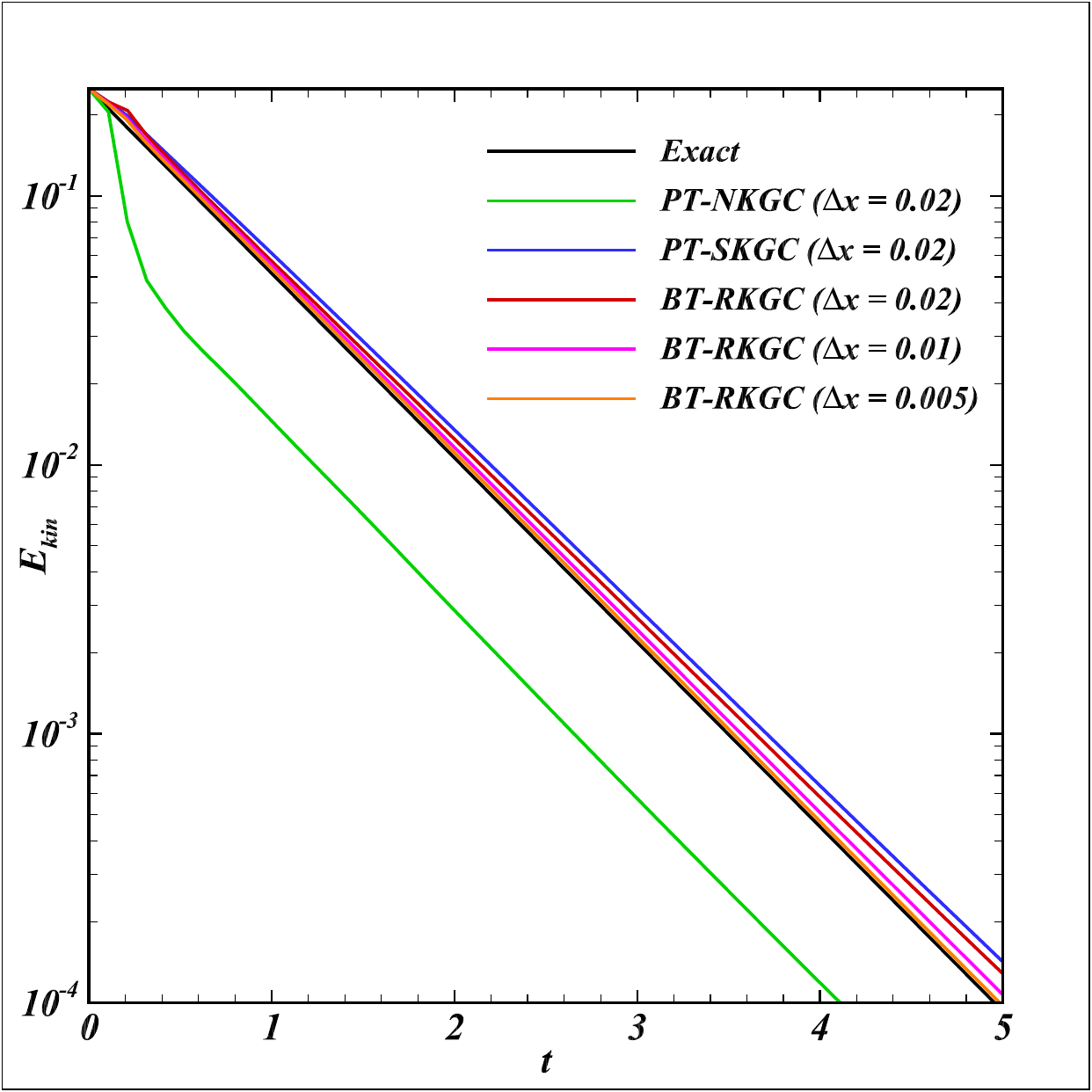}
			\label{taylorgreentransportvelocityTKE}}
		\subfigure[]{\includegraphics[width=.5\textwidth]
			{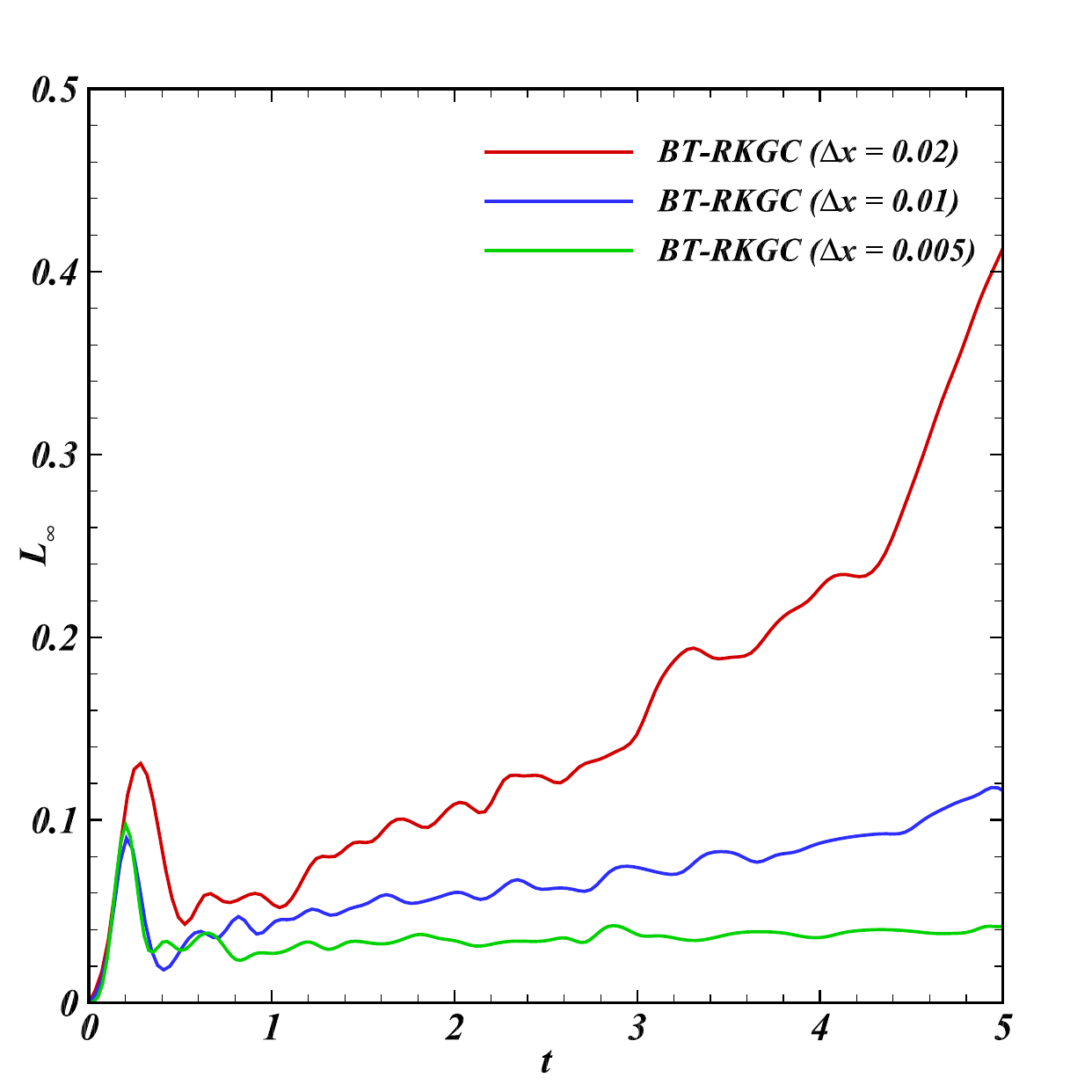}
			\label{taylorgreentransportvelocityTKEERROR}}}
	\caption{Taylor-Green Vortex: WCSPH results obtained with 
		the transport-velocity formulations. (a) Decay of the 
		kinetic energy; (b) Relative error of the maximum 
		velocity (BT-RKGC).}
	\label{taylorgreentransportvelocity}
\end{figure}

The Taylor-Green vortex problem is first investigated using the 
WCSPH method, with the particle spacing $\Delta x =0.02$ 
($50\times50$ particles), $\Delta x=0.01$ ($100\times100$ particles), 
and $\Delta x=0.005$ ($200\times200$ particles).
The initial particle distribution follows a lattice arrangement.
Fig. \ref{taylorgreentransportvelocity} showcases the results 
obtained by employing the transport-velocity formulation in 
each advection time step.
Here, PT and BT denote the original and KGC transport-velocity 
formulations, respectively.
The relative error of the maximum velocity $L_{\infty}^{\rm v}$ 
is defined as
\begin{equation}
	L_{\infty}^{\rm v}=\left|\frac{{\rm max}(\boldsymbol
		{\rm v_{i}}(t))-Ue^{bt}}{Ue^{bt}}\right|.
\end{equation} 
Fig. \ref{taylorgreentransportvelocityTKE} shows that the 
corrected formulations deliver superior results compared to 
PT-NKGC, where no correction is employed. 
Specifically, BT-RKGC yields the most favorable outcomes, 
indicating improved accuracy.
Moreover, as the resolution increases, the kinetic energy obtained 
by BT-RKGC converges toward the analytical solutions.
Such convergence is also demonstrated in Fig. 
\ref{taylorgreentransportvelocityTKEERROR} through the relative error 
of the maximum velocity.
\subsubsection{ESPH results}
\begin{figure}[htb!]
	\centering
	\makebox[\textwidth][c]{
		\subfigure[]{\includegraphics[width=.5\textwidth]
		{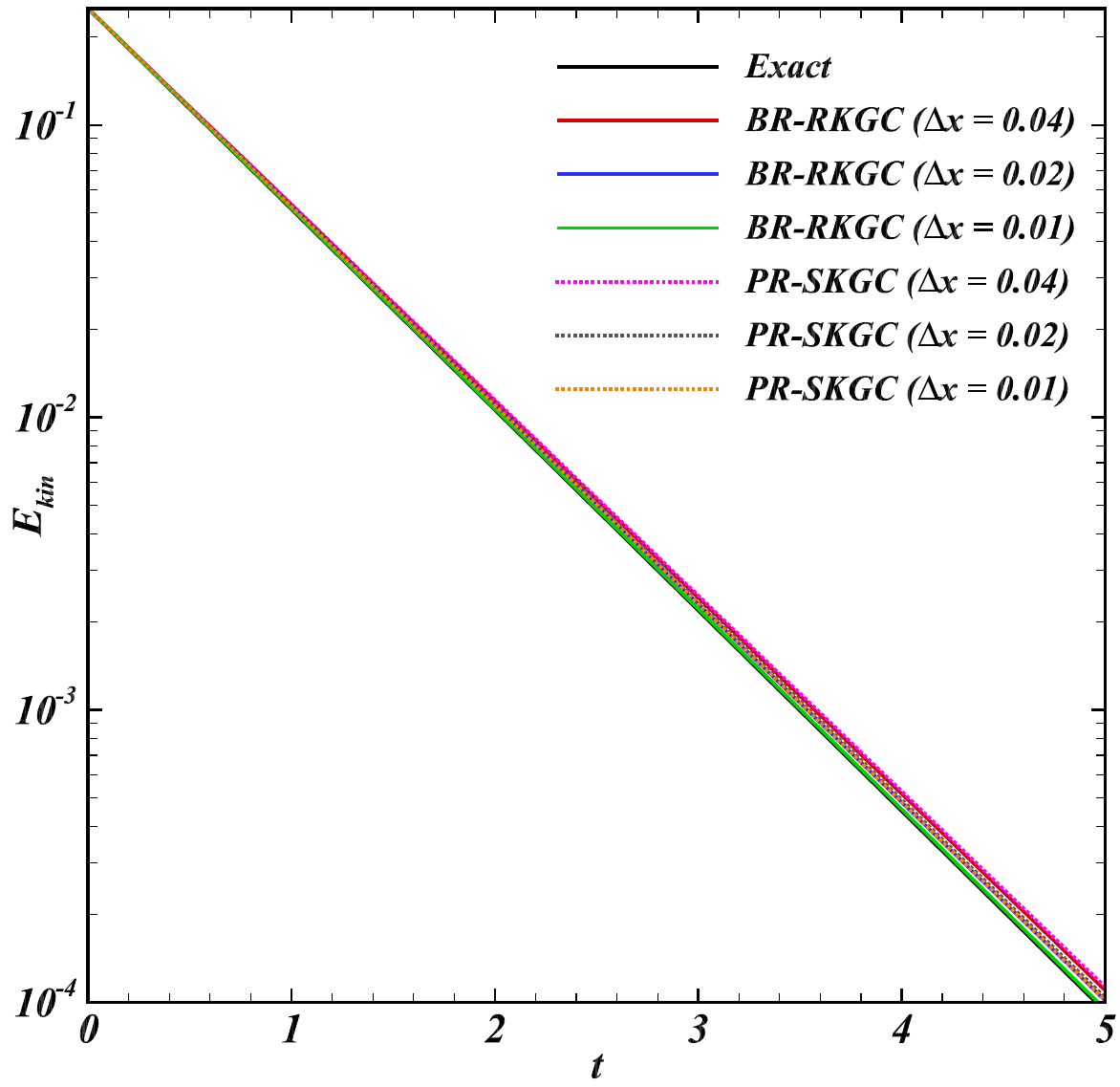}
		\label{taylorgreeneulerianTKE}}
		\subfigure[]{\includegraphics[width=.5\textwidth]
		{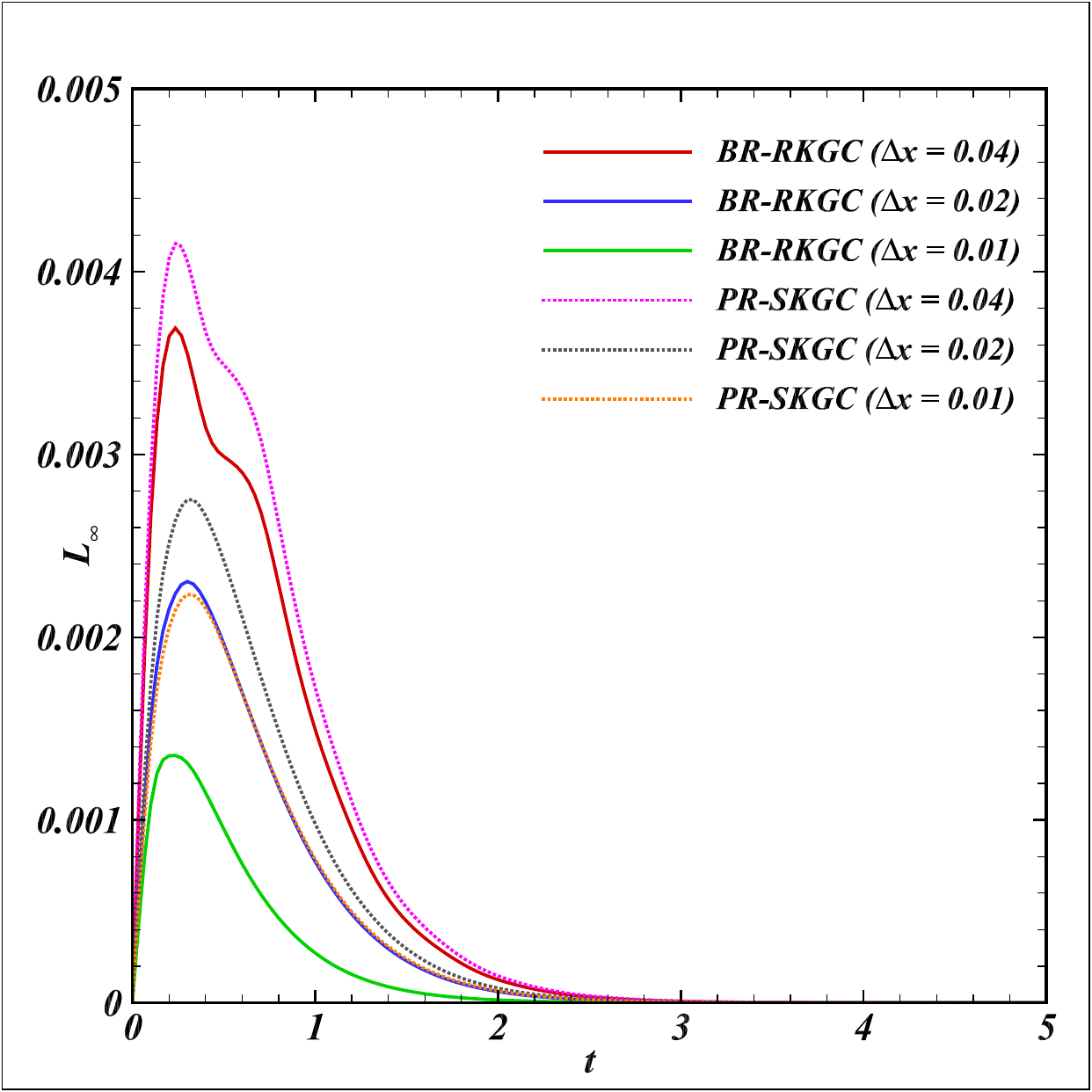}
		\label{taylorgreeneulerianTKEerror}}}
	\caption{Taylor-Green Vortex: ESPH results obtained 
		with the relaxed particles.
		(a) Decay of the kinetic energy;
		(b) Relative error of the kinetic energy.} 
	\label{taylorgreeneulerianTKEanalysis}
\end{figure}

In the ESPH simulations, the particle distributions are initialized 
through relaxation, ensuring that the zero-order consistency residue 
diminishes to less than $1\times10^{-5}$ for P and B relaxations in 
SKGC and RKGC, respectively. 
The initial particle spacing is varied as $\Delta x =0.04$ 
($25\times25$ particles), $\Delta x=0.02$ ($50\times50$ particles), and 
$\Delta x=0.01$ ($100\times100$ particles) to analyze the influence 
of the resolution. 
Fig. \ref{taylorgreeneulerianTKEanalysis} presents the decay of 
the kinetic energy $E_{\text{kin}}$ and its relative error 
$L_{\infty}^{E}$ defined by
\begin{equation}
	L_{\infty}^{E}=\left|\frac{E_{\text{kin}}(t)-
	E_{\text{kin}}^{a}(t)}{E_{\text{kin}}^{a}(t)}\right|,
\end{equation}
where $E_{\text{kin}}^{a}$ represents the analytical kinetic energy 
with a decay rate of $-16\pi^{2}/Re$.
Both SKGC and RKGC produce results converging to the analytical solution, 
while RKGC achieves lower kinetic energy error across all resolutions and 
exhibits improved accuracy compared to that of SKGC.
\subsection{Lid-driven cavity at Re=1000}
The lid-driven cavity problem serves as a well-known and 
challenging test case for the SPH method.
In this scenario, a wall-bounded unit square cavity is presented, 
with its top wall moving at a constant speed of $U_{\text{wall}}=1$.
For the flow at a Reynolds number of $Re = 1000$, we reference 
the high-resolution multi-grid results of Ghia et al. \cite{ghia1982high}, 
who utilized the finite difference method on a $257\times257$ mesh.
\subsubsection{WCSPH results}
\begin{figure}[htb!]
	\centering
	\vspace{-3mm}
	\makebox[\textwidth][c]{
		\subfigure[]{\includegraphics[width=.5\textwidth]
		{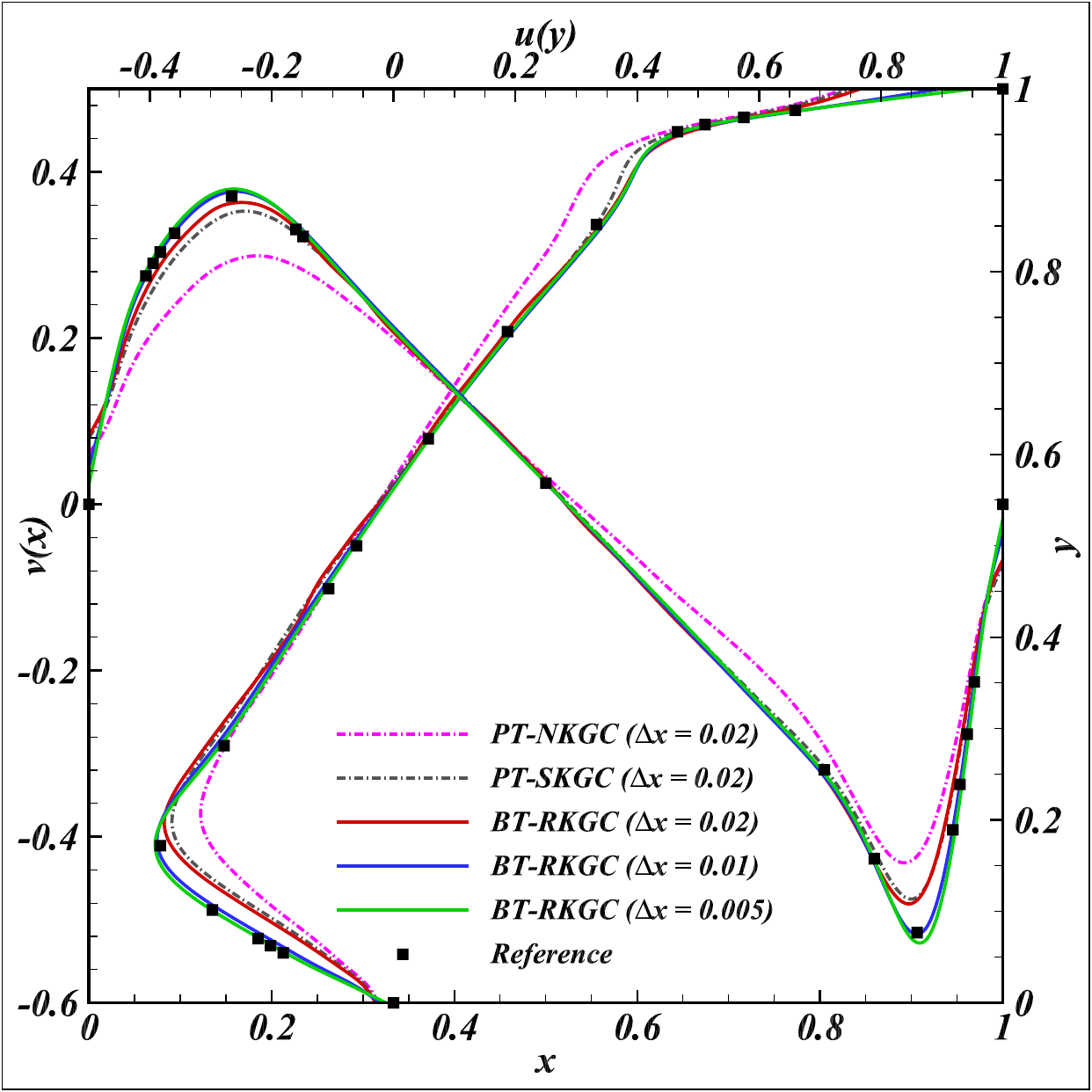}
		\label{liddrivensquarecavitycomH13}}
		\subfigure[]{\includegraphics[width=.5\textwidth]
		{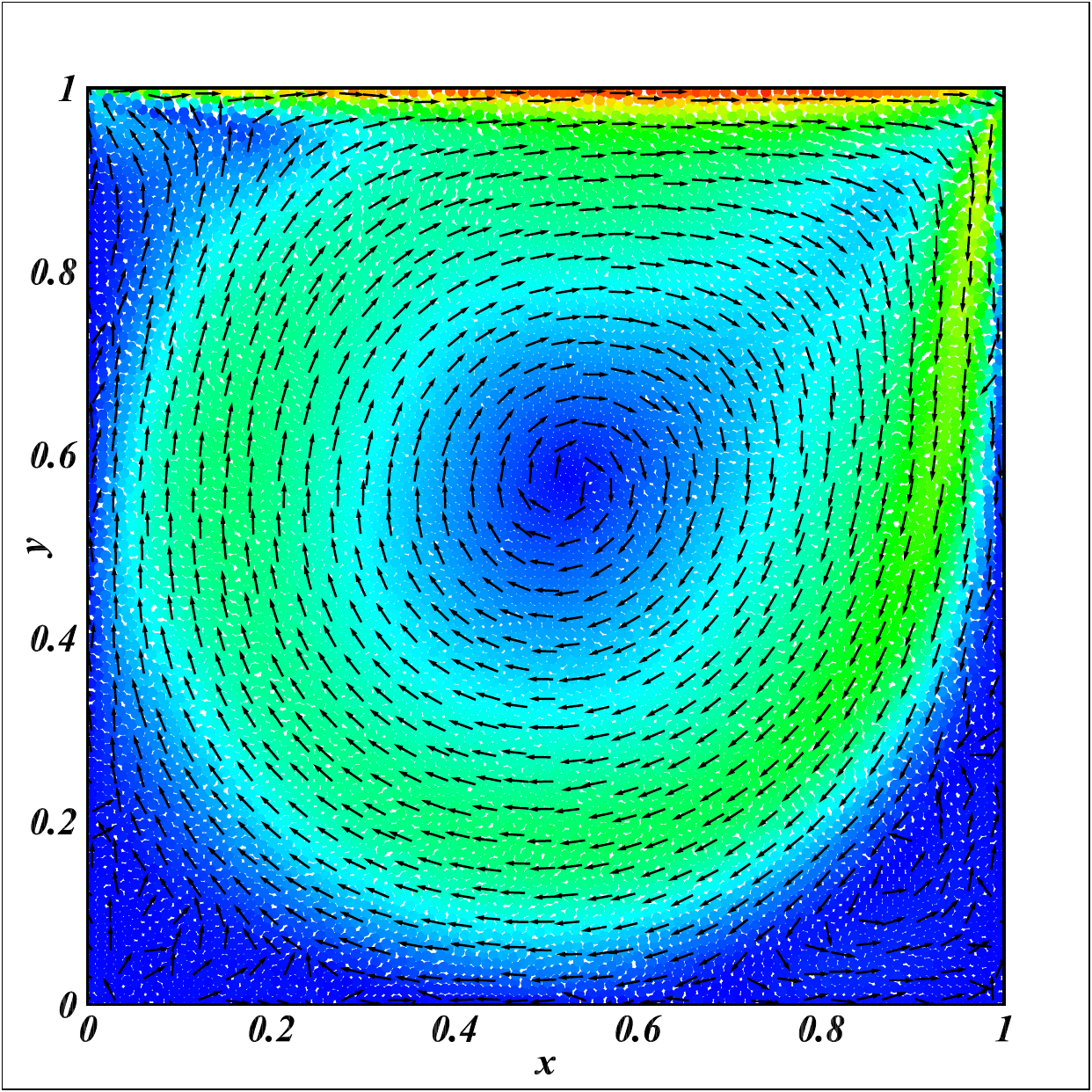}
		\label{liddrivensquarecavityH13field}}}
	\caption{Lid-driven cavity: WCSPH results obtained with the 
		transport-velocity formulations. 
		(a) Velocity profiles along the horizontal and vertical 
		central lines; (b) Velocity fields (visualized by the 
		magnitude ranging from 0 to 1) with vectors obtained by 
		BT-RKGC at the $\Delta x=0.01$.}
	\label{liddrivensquarecavityWCSPH}
\end{figure}

The results obtained by the WCSPH method are presented in 
Fig. \ref{liddrivensquarecavityWCSPH}.
The comparison of velocity profiles with the reference results, 
depicted in Fig. \ref{liddrivensquarecavitycomH13}, reveals that 
corrected formulations yield results more closely aligned with the 
reference than PT-NKGC formulation without correction.
Furthermore, BT-RKGC outperforms PT-SKGC.
As resolution increases, results obtained by BT-RKGC demonstrate 
clear convergence and good agreement with the reference.
The magnitude of the velocity and the velocity vectors of the flow 
field shown in Fig. \ref{liddrivensquarecavityH13field} exhibit a 
smoothed distribution and typical vortical structures, such as those 
induced by the shear force of the moving wall and the single-core 
vortex located at the center of the cavity, consistent with findings 
presented in Refs. \cite{ghia1982high, adami2013transport, 
	zhu2021consistency}.
\subsubsection{ESPH results}
\begin{figure}[htb!]
	\centering
	\makebox[\textwidth][c]{
		\subfigure[]{\includegraphics[width=.5\textwidth]
			{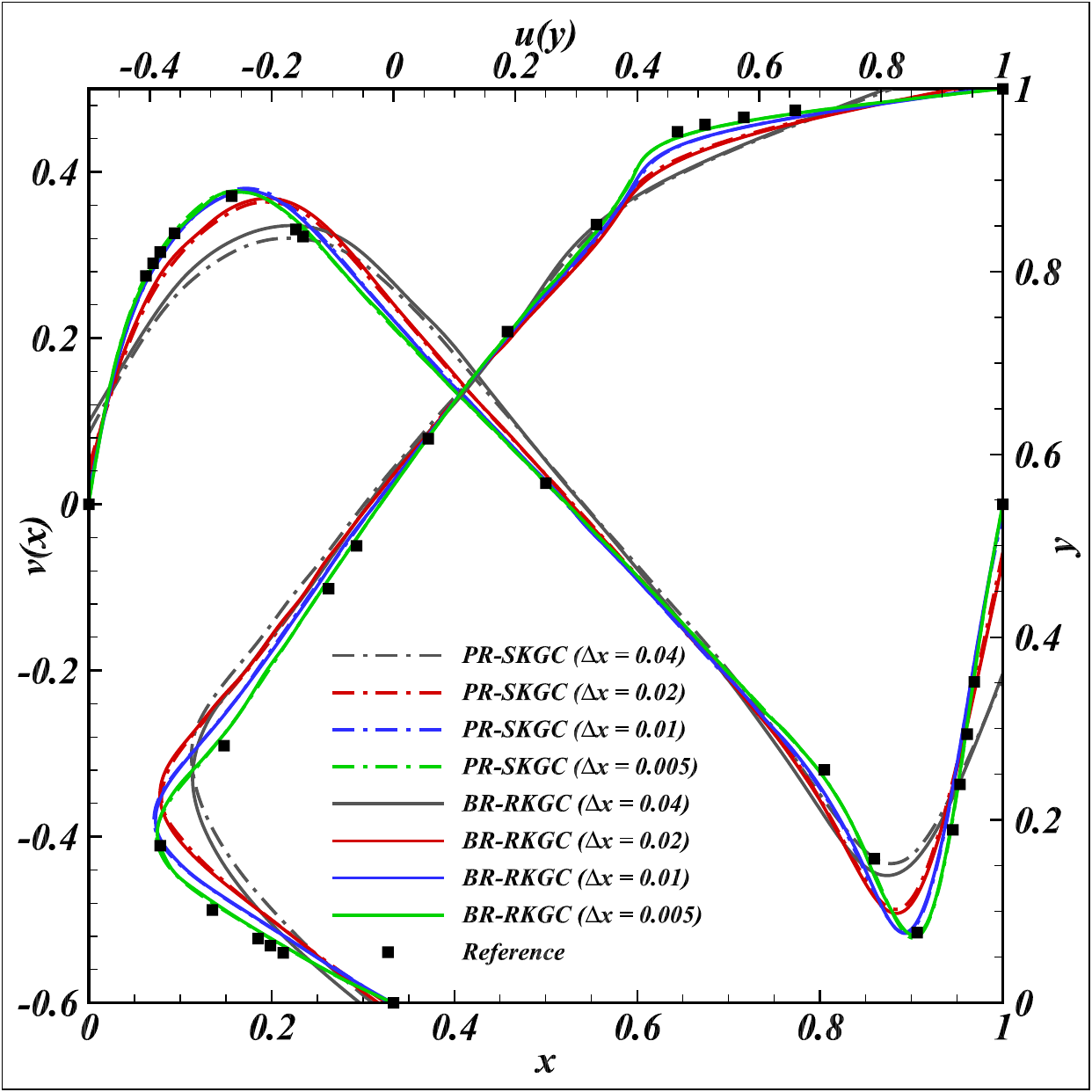}
		\label{liddrivensquarecavitH13}}
		\subfigure[]{\includegraphics[width=.5\textwidth]
			{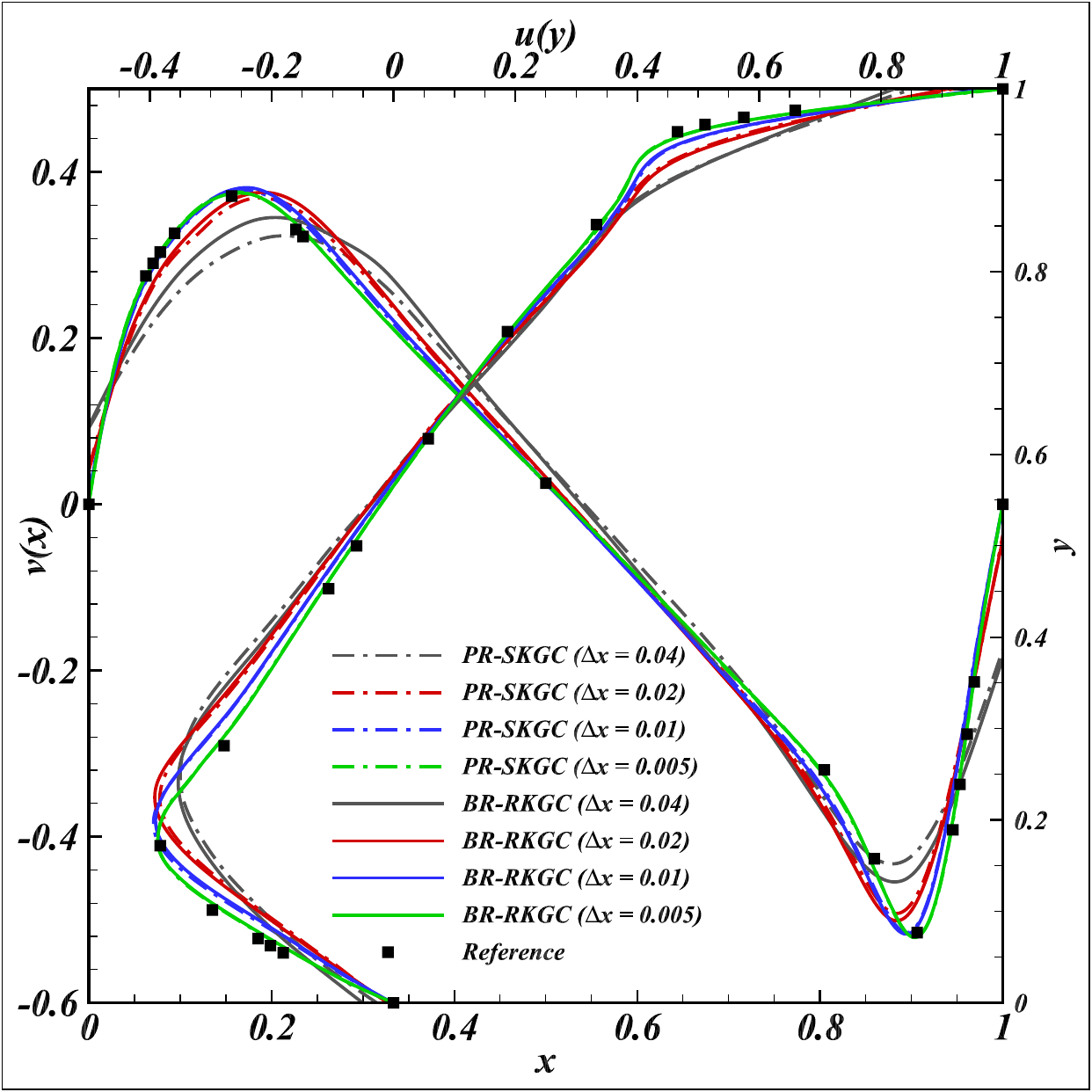}
		\label{liddrivensquarecavitH115}}}
	\caption{Lid-driven cavity: ESPH method results for velocity 
		profiles along the horizontal and vertical central lines 
		at different smoothing lengths. 
		(a) $h=1.3\Delta x$; (b) $h=1.15\Delta x$.}
	\label{liddrivensquarecavityvelovityH13H115}
\end{figure}

Again, the ESPH results are obtained on fully relaxed initial particle 
distributions at smoothing lengths of $h=1.3\Delta x$ and $h=1.15\Delta x$, 
as illustrated in Fig. \ref{liddrivensquarecavityvelovityH13H115}..
It is observed that the BR-RKGC produces results closer to the reference than 
the PR-SKGC at the low resolutions, i.e., $\Delta x=0.04$ and $\Delta x=0.02$, 
indicating less integration errors.
With the resolution increased, both methods could yield converged results, 
and the error difference is small due to the sufficient smoothing length.

\begin{figure}[htb!]
	\centering
	\makebox[\textwidth][c]{\subfigure[]{
			\includegraphics[width=.5\textwidth]
			{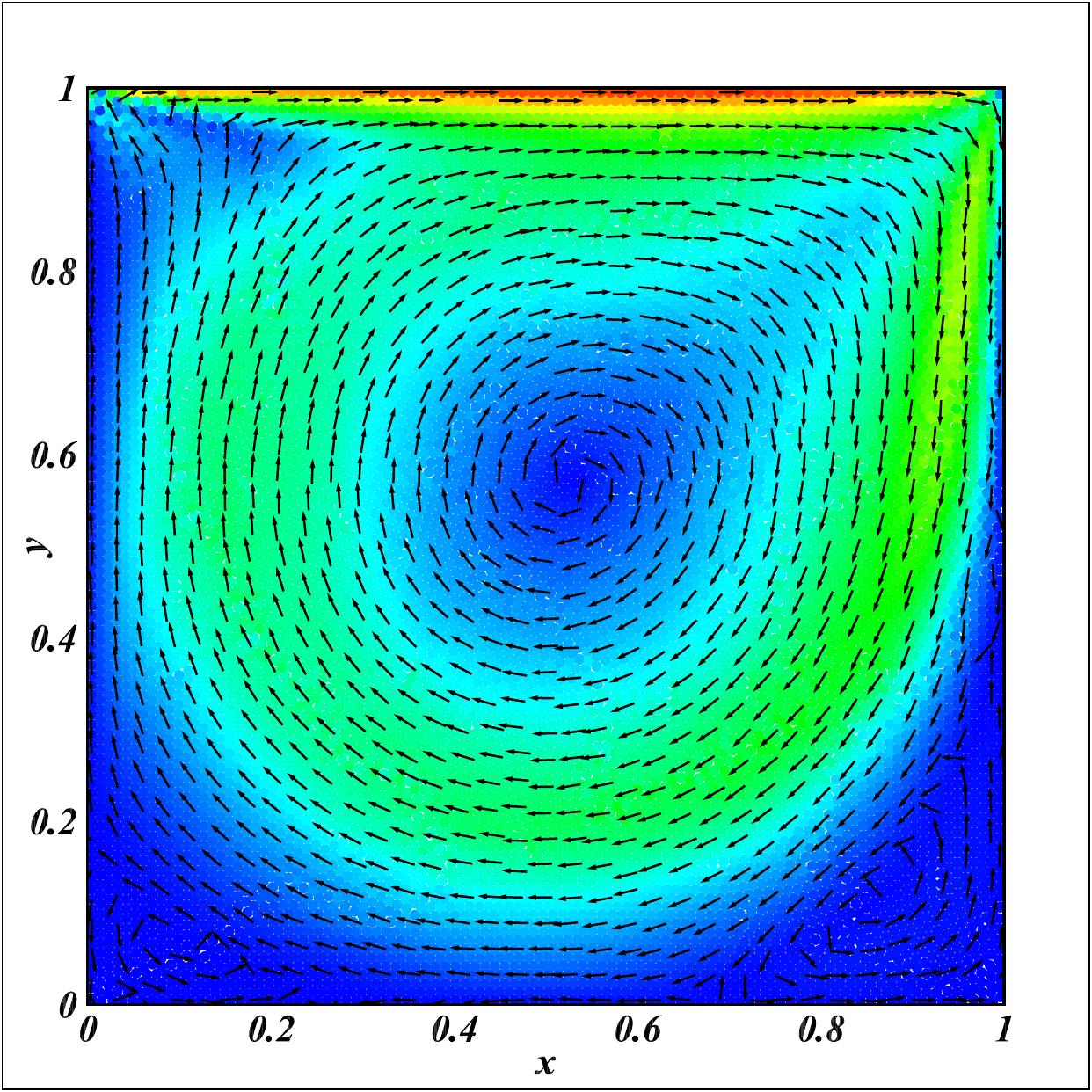}
			\label{liddrivensquarecavitH08RKGC}}
		\subfigure[]{
			\includegraphics[width=.5\textwidth]
			{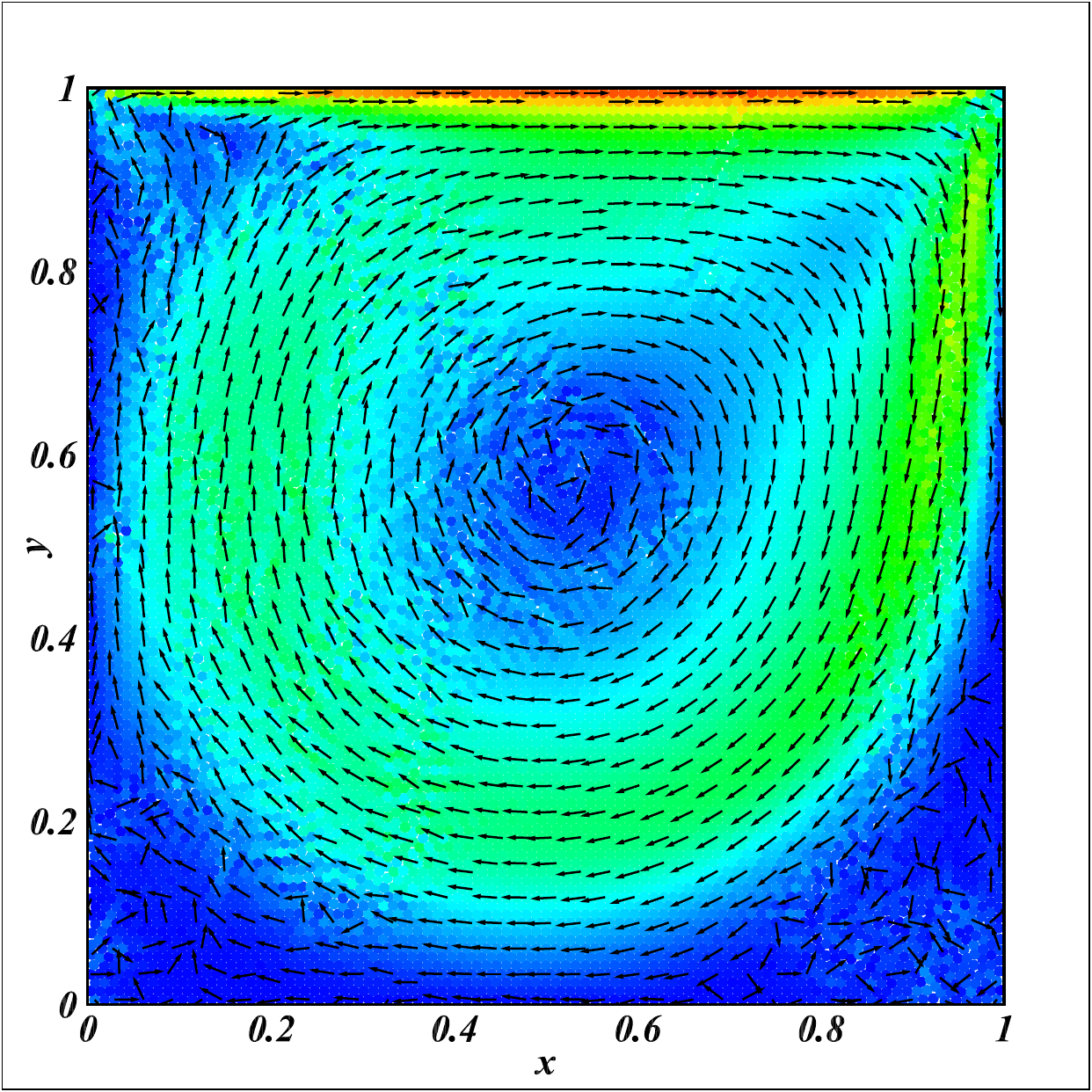}
			\label{liddrivensquarecavitH08SKGC}}}
	\caption{Lid-driven cavity: ESPH method results for velocity 
		fields (visualized by the magnitude ranging from 0 to 1) 
		with vectors at $h=0.8\Delta x$ and $\Delta x=0.01$. 
		(a) BR-RKGC; (b) PR-SKGC.}
	\label{liddrivensquarecavityvelovityH08}
\end{figure}

Fig. \ref{liddrivensquarecavityvelovityH08} presents the obtained velocity 
fields with vectors when the smoothing length is decrease to $h=0.8\Delta x$.
As discussed in Section \ref{convergenceanalysis}, the RKGC formulation 
combined with the B relaxation achieves the first-order consistency, not 
explicitly relying on the smoothing length, while the SKGC one suffers 
serious degeneration.
Therefore, as presented in Fig. \ref{liddrivensquarecavitH08RKGC}, the 
BR-RKGC is still able to generate the smooth velocity distribution and 
captures the key flow characteristics, aligning with the reference 
results \cite{ghia1982high, bayareh2019explicit} and the one displayed in 
Fig. \ref{liddrivensquarecavityH13field} obtained by the WCSPH method.
Moreover, the secondary vortices in the lower and upper corners are still 
well-identified.
However, the PR-SKGC, as presented in Fig. \ref{liddrivensquarecavitH08SKGC}, 
fails to yield a reasonable smooth velocity distribution, with velocity 
oscillation and noise, especially at the upper-left corner vortex and 
around the single-core vortex. 
The velocity vectors at the lower corners also failed to capture the 
secondary vortices.
Therefore, the proposed RKGC formulation and B relaxation could still 
achieve improved accuracy and good convergence, even with a reduced 
$h/\Delta x$ value.
Note that, in practical viscous flow simulations, while the RKGC aims to 
achieve first-order consistency for gradient and divergence operators, 
the numerical results do not necessarily always maintain it because the 
Laplacian operator in SPH approximations does not precisely satisfy 
first-order consistency yet, and the adoption of the Riemann solver may 
also degenerate the consistency as the RKGC is only applied on the 
non-dissipative terms.
\subsection{Standing wave}
\begin{figure}[htbp]
	\centering
	\includegraphics[width=0.75\textwidth]
	{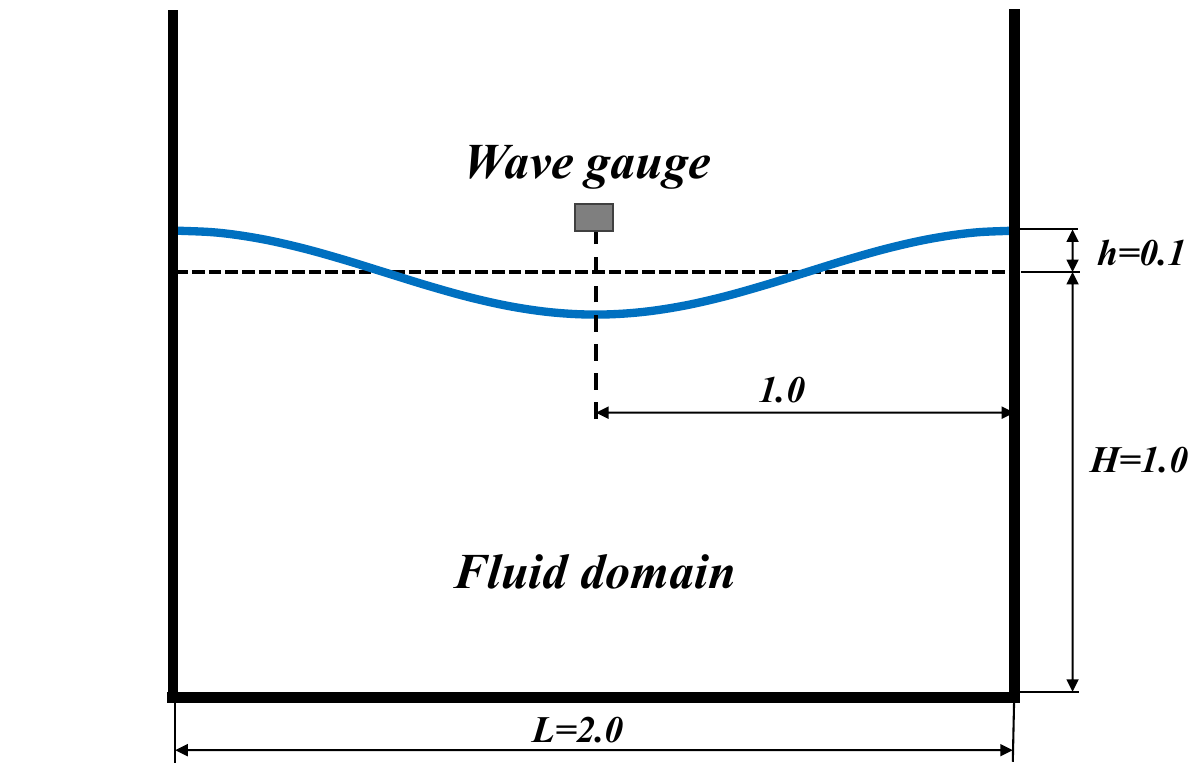}
	\caption{Standing wave: Initial 
		configuration of the simulation.}
	\label{standingwaveconfiguration}	
\end{figure}

The standing wave problem serves as a typical benchmark for 
evaluating the accuracy of the SPH method in addressing 
free-surface problems.
In this section, a two-dimensional standing wave problem is 
investigated using the WCSPH method.
The initial configuration, as depicted in Fig. 
\ref{standingwaveconfiguration}, 
defines the initial free surface according to
\begin{equation}
	\eta_{0} = A \cos(k(x+\lambda)/2).
\end{equation}
Here, the parameters are set as follows: the wave amplitude
$A$ = $0.1H$, with an average water depth of $H=1.0$;
the wave number $k =2 \pi/\lambda$, and the wavelength 
$\lambda =2.0$.
The initial velocity of the particles is set to be zero.
Note that transport velocity formulations are not employed here 
(also for other free-surface flows in this work) following the 
general practice of SPH simulations \cite{oger2007improved, 
	lyu2023derivation, ren2023efficient}.
The study evaluates the free-surface elevation at the central 
position, and compare it against the second-order analytical 
solution proposed by Wu et al. \cite{wu1994finite}.
In order to maintain numerical stability, the KGC matrix near 
the free surface is weighted by the identity matrix (WKGC$^1$), 
as suggested in Ref. \cite{ren2023efficient}, with the constant 
parameter $\alpha = 0.5$ for all free-surface flow problems in 
the current study. 

\begin{figure}[htb!]
	\centering
	\includegraphics[width=\textwidth]
		{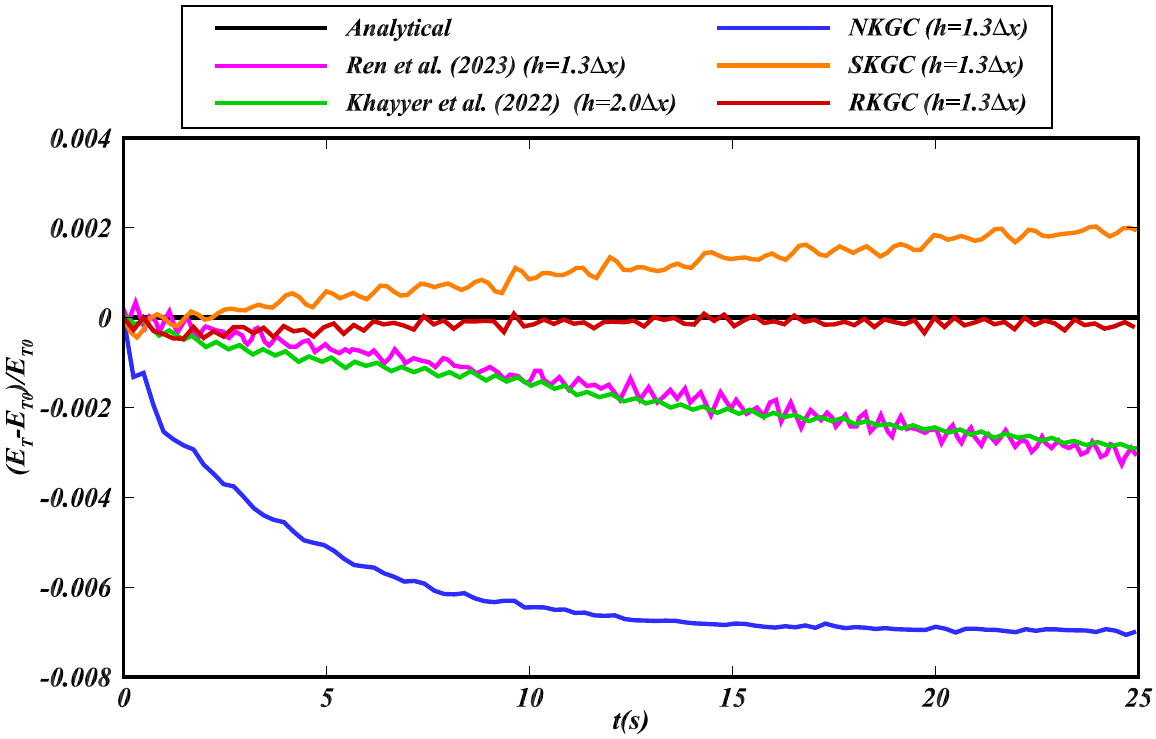}
	\caption{Standing wave: Time evolution of the decay of the 
		normalized mechanical energy obtained by different 
		formulations ($\Delta x=0.01$).}
	\label{standingwaveenergy}	
\end{figure}

Fig. \ref{standingwaveenergy} illustrates the decay of the normalized 
mechanical energy across different formulations and compares them 
with the analytical and reference results.
It is observed that the RKGC formulation is able to preserve the 
energy very well, suggesting very small numerical dissipation.
However, NKGC exhibits rapid energy decay, even when the smoothing 
length is increased to $h=2.0\Delta x$, as shown by Khayyer et al. 
\cite{khayyer2023enhanced}.
It is also noted that the SKGC formulation leads to an increase in 
the energy, consistent with findings from Ref. 
\cite{ren2023efficient, lyu2023derivation}.
Therefore, extra weight with the identity matrix (as WKGC$^2$ in 
Ref. \cite{ren2023efficient}) is added to decrease the contribution 
of the SKGC formulation to eliminate the artifact but, as illustrated 
in Fig. \ref{standingwaveenergy}, it still shows considerable 
energy loss.

\begin{figure}[htbp]
	\centering
	\vspace{-2.5cm}
	\subfigure[]{\includegraphics[width=\textwidth]
		{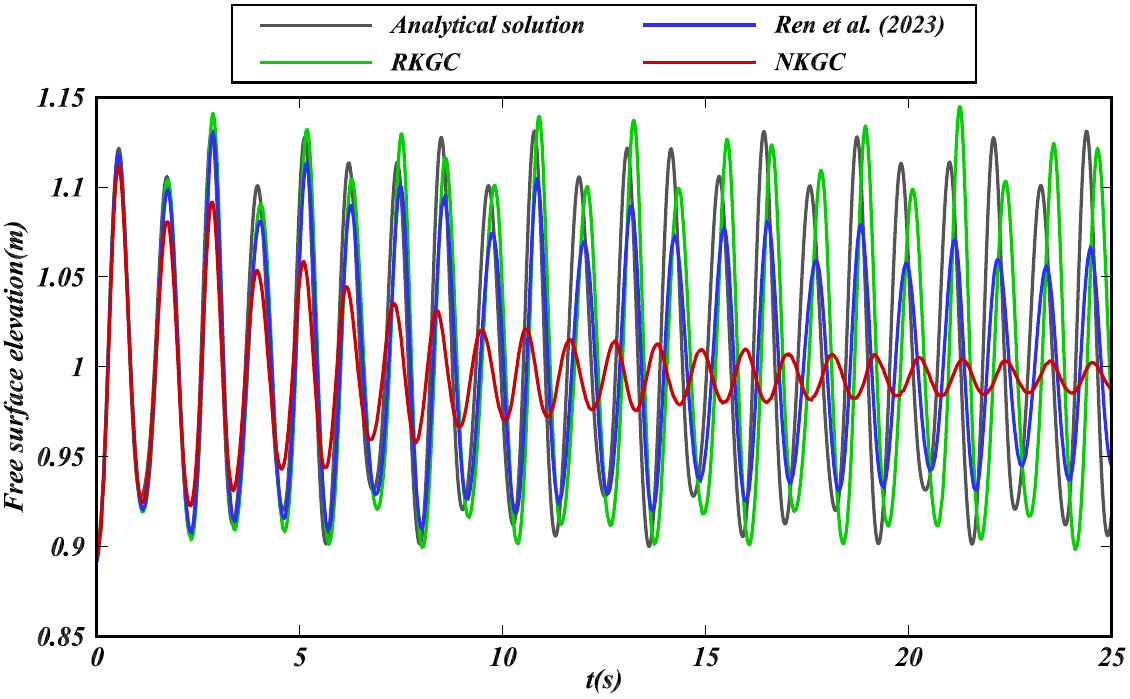}
		\label{standingwavewaveheightschemes}}
	\subfigure[]{\includegraphics[width=\textwidth]
		{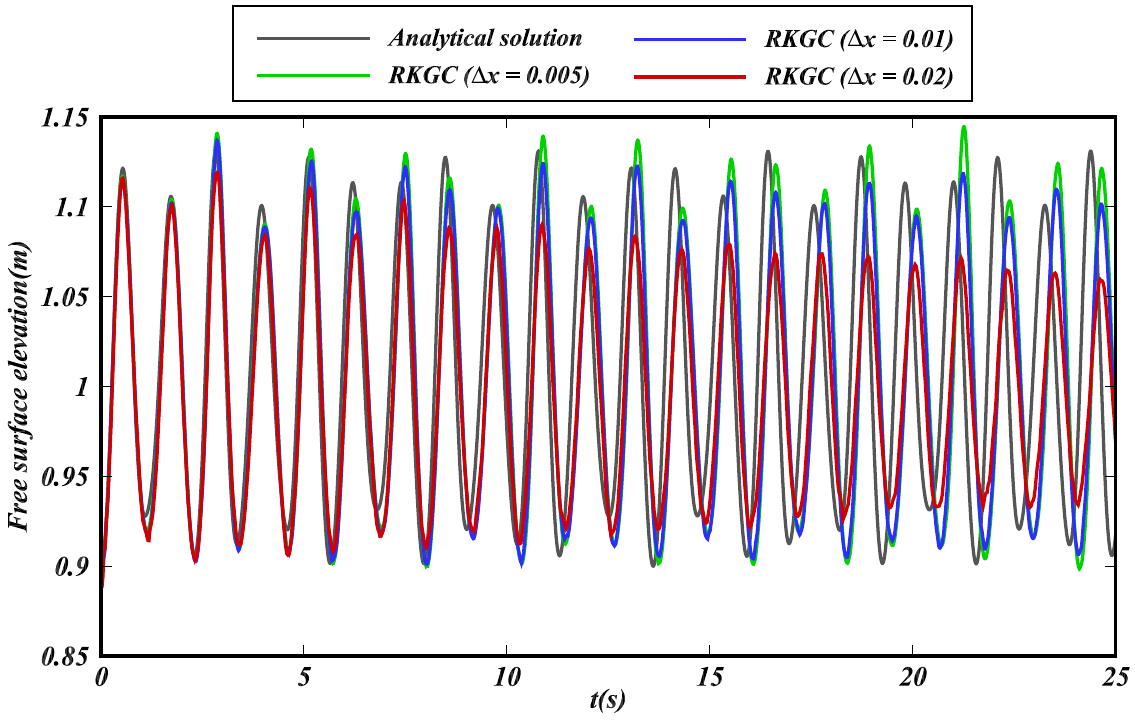}
		\label{standingwavewaveheightresolution}}
	\caption{Standing wave: Time evolution of the 
		free-surface elevation at the center of the tank.
		(a) Comparison across different formulations 
		($\Delta x = 0.005$);
		(b) Convergence study of the RKGC formulation.}
	\label{standingwavewaveheight}
\end{figure}

Fig. \ref{standingwavewaveheight} illustrates the wave heights across 
different methods as well as the convergence analysis of the RKGC 
formulation.
The comparisons of the wave height depicted in Fig. 
\ref{standingwavewaveheightschemes} indicate that while the results 
obtained with both SKGC and RKGC achieve notable improvements 
compared to NKGC, RKGC further improves accuracy considerably 
and generates closer approximations to the analytical solution.
As displayed in Fig. \ref{standingwavewaveheightresolution}, RKGC 
demonstrates good convergence with increasing resolution.
\subsection{Oscillating drop}
\begin{figure}[htbp]
	\centering
	\includegraphics[width=0.5\textwidth]
		{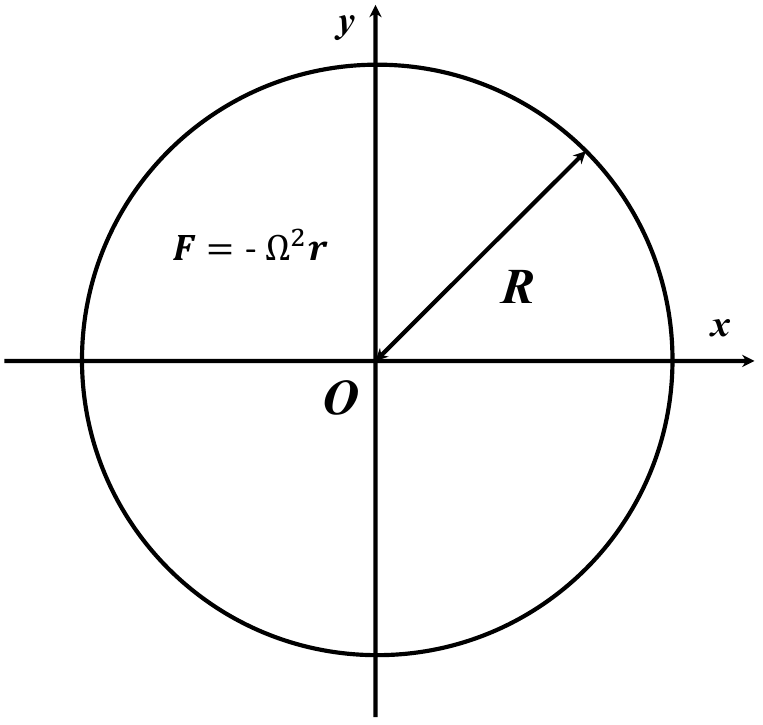}
	\caption{Oscillating drop: Schematic 
		illustration of the benchmark test.}
	\label{oscialltingdropconfiguration}	
\end{figure}

The two-dimensional oscillating drop was also investigated to evaluate 
the energy conservation properties of the proposed method.
This problem, as outlined in Ref. \cite{antuono2015energy} and depicted 
in Fig. \ref{oscialltingdropconfiguration}, involves a drop with a  
radius of $R=1$ immersed in an assumed inviscid fluid.
This drop experiences a central conservative force $F=-\Omega^{2}R$ 
and is initialized with a velocity profile defined by

\begin{equation}
	\begin{cases}
		u_{0}=A_{0}x \\
		v_{0}=-A_{0}y
	\end{cases},
\end{equation}
where $A_{0}=1.0$ and $A_{0}/\Omega=1.0$. 
The analytical solution reported in Ref. \cite{monaghan2013simple} is 
referenced for quantitative comparison and validation.

\begin{figure}[htbp]
	\centering
	\makebox[\textwidth][c]{\subfigure[]{
			\includegraphics[width=.5\textwidth]
			{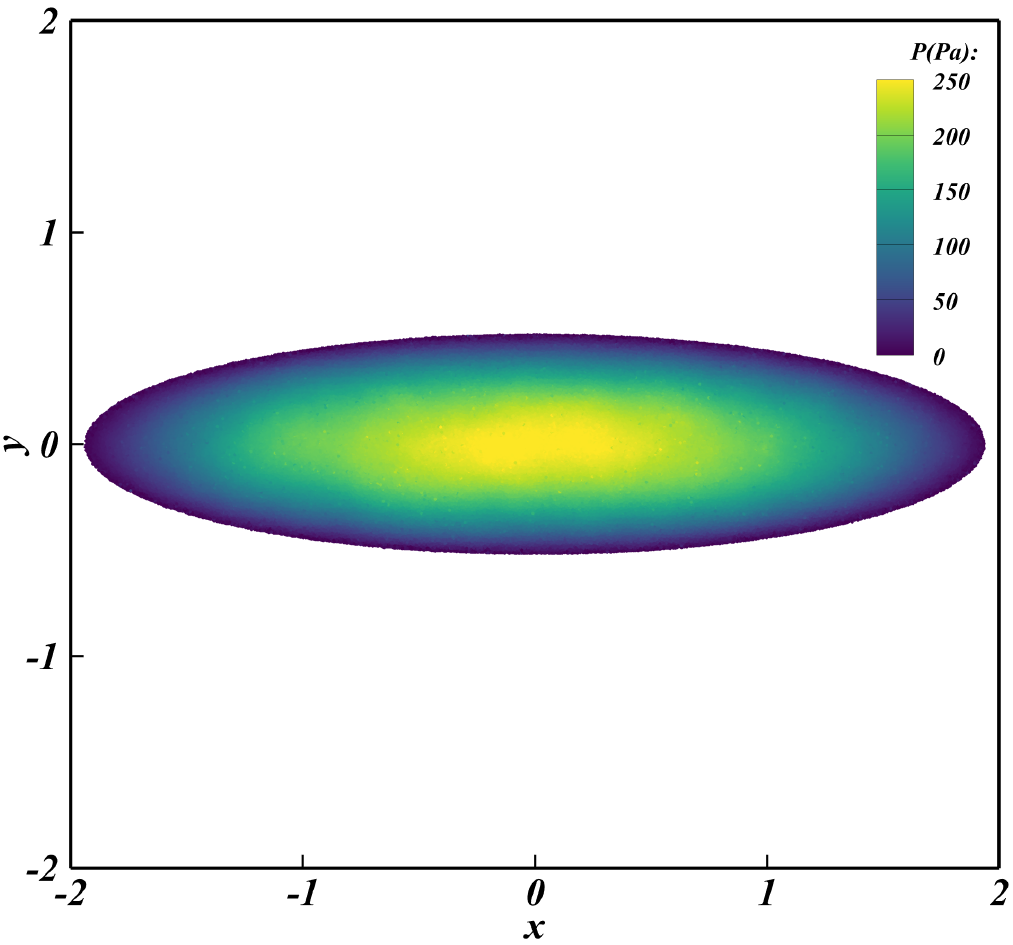}
			\label{oscialltiondroppressurefieldtime1}}
		\subfigure[]{
			\includegraphics[width=.5\textwidth]
			{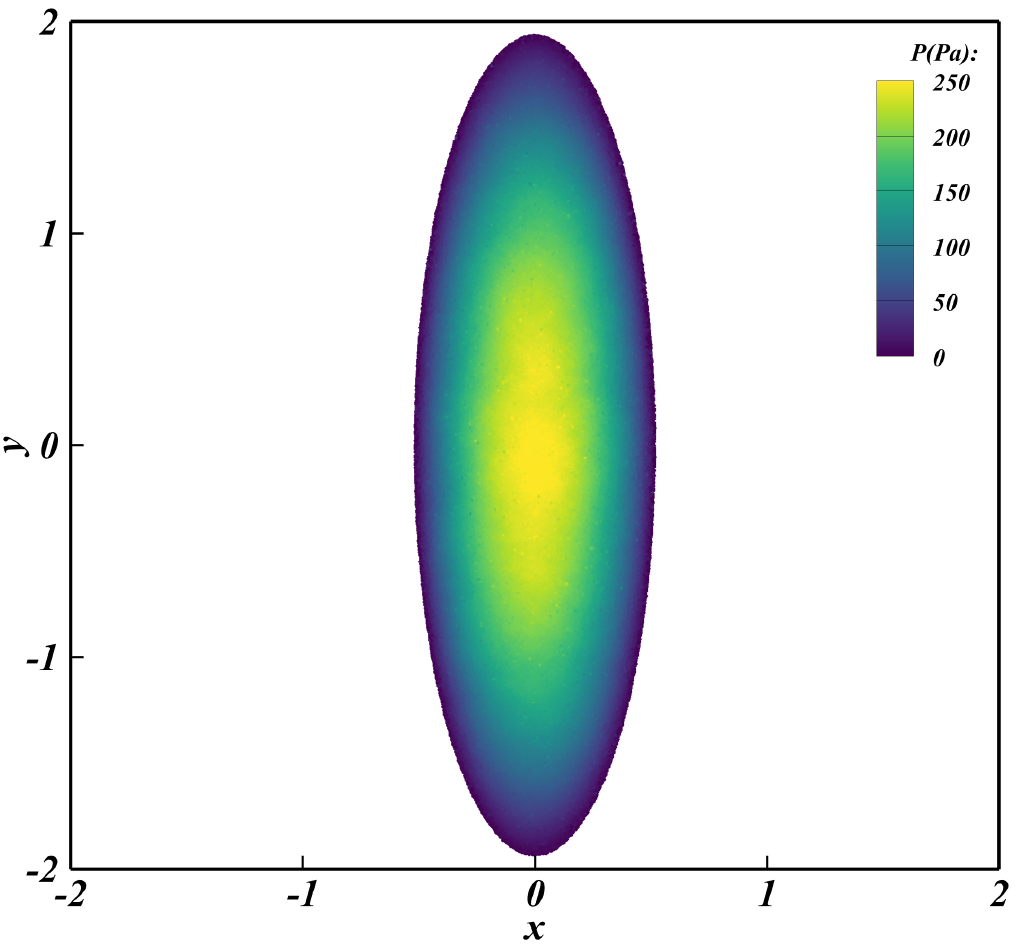}
			\label{oscialltiondroppressurefieldtime2}}}
	\caption{Oscillation drop: Snapshots of the free-surface 
		profile and the pressure contour obtained by the RKGC. 
		(a) t=20.5s; (b)t=22.9s.}
	\label{oscialltiondroppressurefield}
\end{figure}

\begin{figure}[htb]
	\centering
	\includegraphics[width=\textwidth]{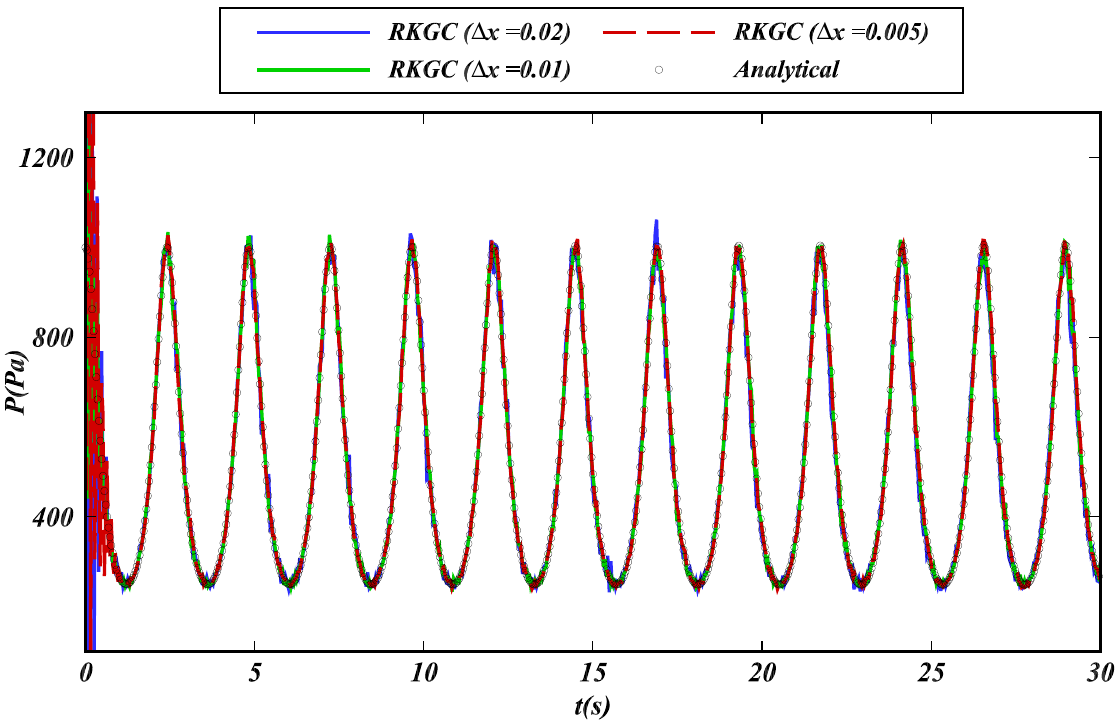}
	\caption{Oscillating drop: Time history of the pressure at the drop 
		center obtained by the RKGC with different particle resolutions.}
	\label{oscialltiondroppressure}	
\end{figure}

The pressure contours at two different instants are illustrated in Fig. 
\ref{oscialltiondroppressurefield}, showcasing the robust free-surface 
profile and smooth pressure fields obtained by RKGC.
Furthermore, Fig. \ref{oscialltiondroppressure} illustrates pressure 
observations at the center of the domain obtained by the RKGC for 
different resolutions.
It shows that the RKGC has good numerical stability and accuracy.
As the resolutions increase, they converge and indicate a good 
convergence property.

\begin{figure}[htb]
	\centering
	\includegraphics[width=\textwidth]{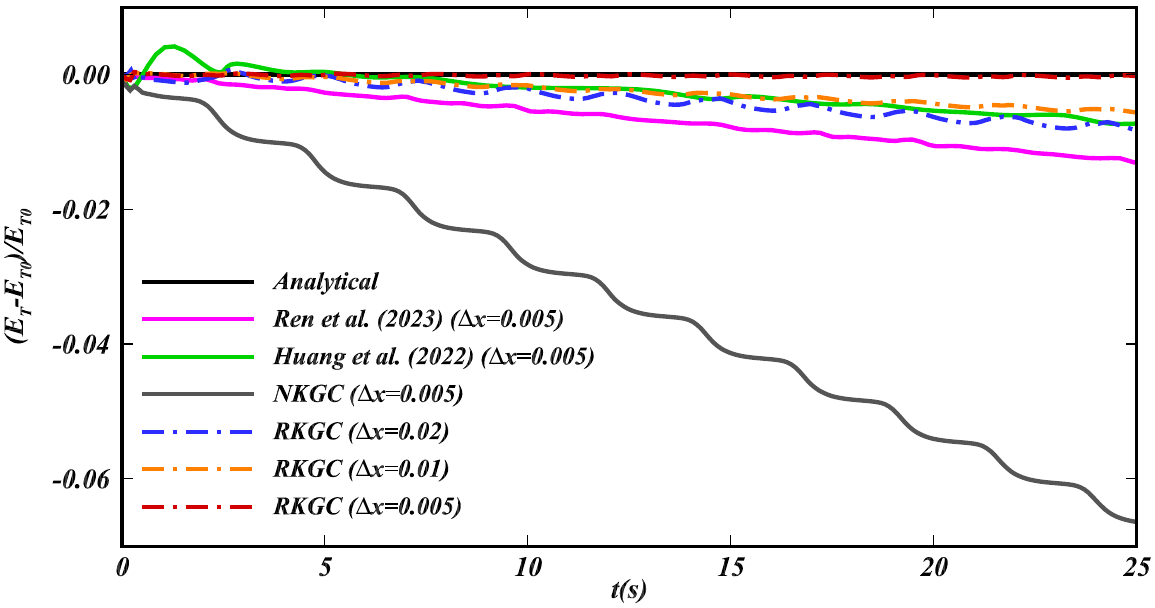}
	\caption{Oscillating drop: Time evolution of the decay of 
		the normalized mechanical energy obtained by different 
		formulations.}
	\label{oscialltiondropenergy}	
\end{figure}

\begin{figure}[htbp]
	\centering
	\vspace{-2.5cm}
	\subfigure[]{\includegraphics[width=\textwidth]
		{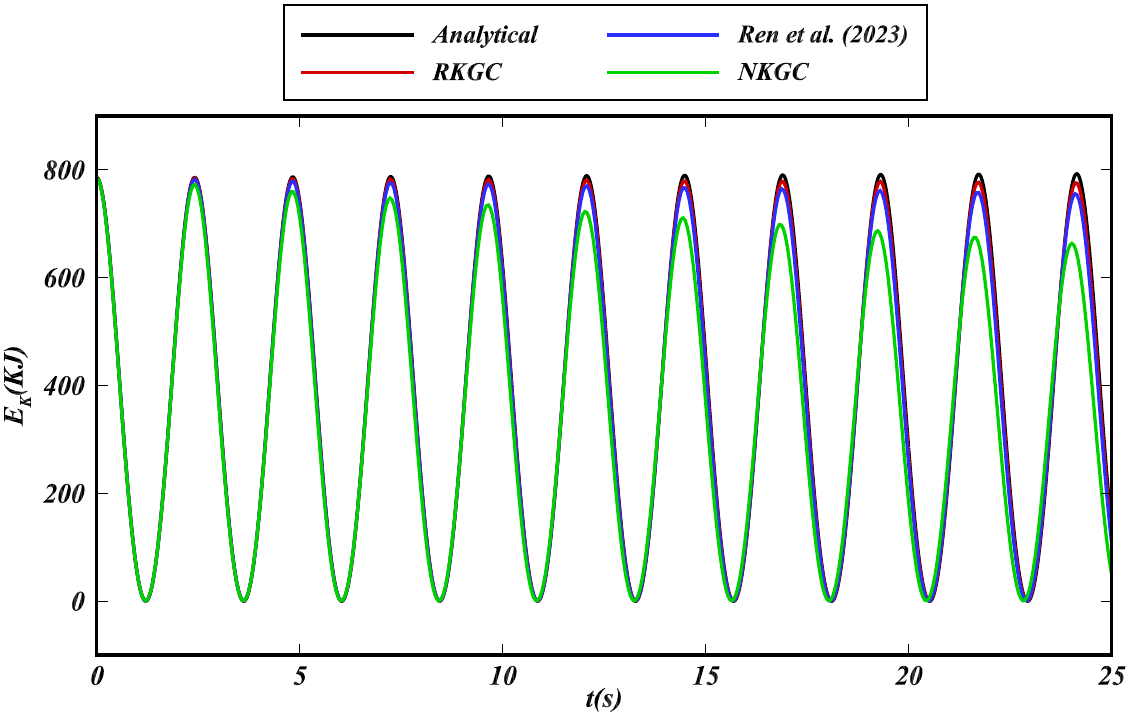}
		\label{oscialltiondropkineticenergyscheme}}
	\subfigure[]{\includegraphics[width=\textwidth]
		{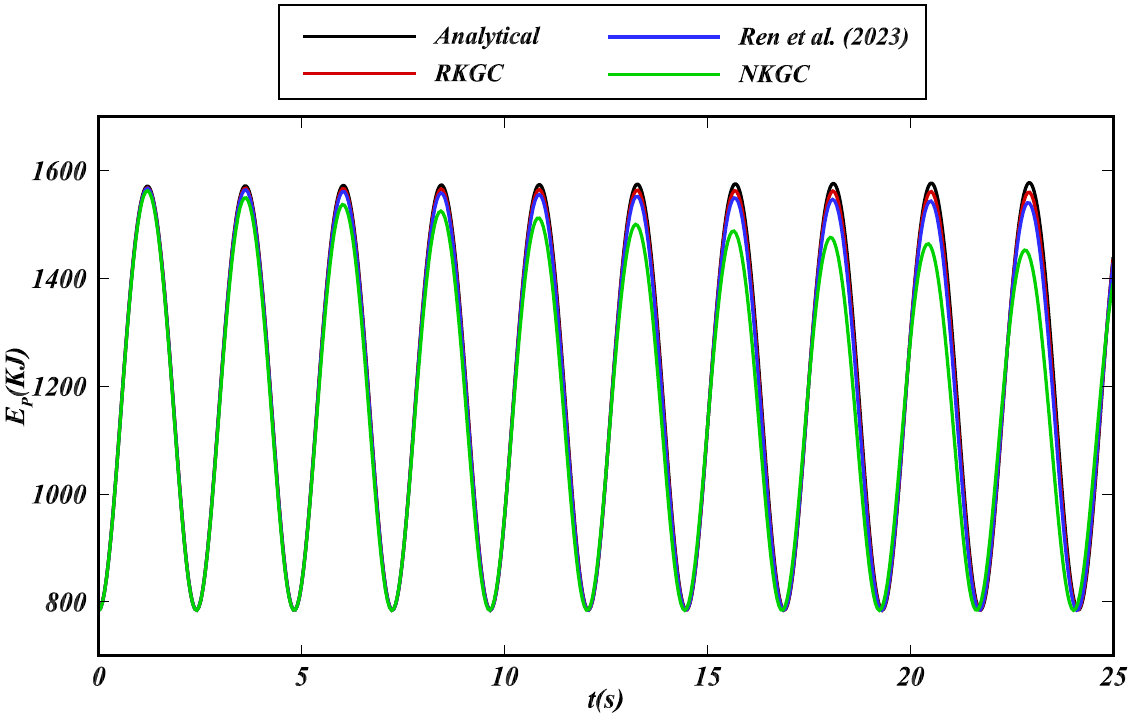}
		\label{oscialltiondroppotentialenergyscheme}}
	\caption{Oscillating drop: Time history of the energy obtained 
		by different formulations ($\Delta x = 0.01$). 
		(a) Kinetic energy; (b) Potential energy.}
	\label{oscialltiondropenergycomparisonscheme}
\end{figure}

Fig. \ref{oscialltiondropenergy} presents the time evolution of the 
decay of normalized mechanical energy across different formulations 
and compares them with the analytical and reference results.
Aligning with the observation in Fig. \ref{standingwaveenergy}, the 
RKGC formulation is able to preserve the energy quite well, even at a 
low resolution, and the energy conservation properties are improved 
with the resolution increase.
It indicates there is no evident energy decay at the resolution of 
$\Delta x = 0.005$.
However, at this high resolution, NKGC still exhibits a considerably 
high decay rate of the energy.
While the corrected formulations introduced by Ren et al. \cite{ren2023efficient} 
and Huang et al. \cite{huang2022development} achieve notable reduction 
of the energy decay, they still suffer more energy loss than that 
obtained by RKGC at a lower resolution.
The separate comparisons of kinetic and potential energies are depicted 
in Fig. \ref{oscialltiondropenergycomparisonscheme}.
The corrected formulations all show improved alignment with the 
analytical solution compared to the NKGC for both kinetic and 
potential energies.
Moreover, RKGC exhibits a closer agreement with the analytical solution
than those reported in Ref. \cite{ren2023efficient} where a 
corrected formulation is also adopted. 
\subsection{Dam-break flow}
\begin{figure}[htb!]
	\centering
	\includegraphics[width=\textwidth]{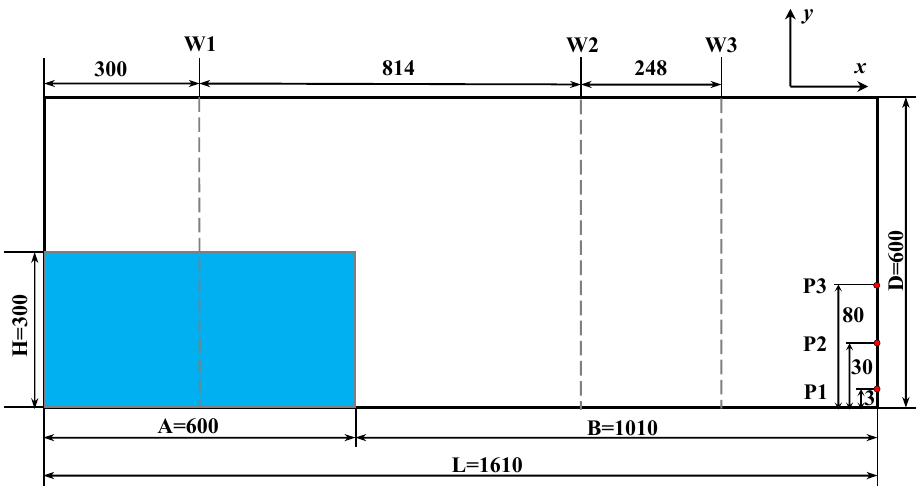}
	\caption{Dam-break flow: Initial configuration of the simulation.}
	\label{dambreakconfiguration}	
\end{figure}

The dam-break flow, extensively explored both experimentally 
\cite{martin1952experimental, buchner2002green, lee2002numerical, 
	lobovsky2014experimental} and numerically 
\cite{ferrari2009new, cercos2015aquagpusph, zhang2017weakly, zhang2020dual}, 
is a challenging benchmark to validate the SPH method.
Fig. \ref{dambreakconfiguration} depicts the initial configuration for the 
simulation, aligning with the experimental setup outlined by Lobovsky et al. 
\cite{lobovsky2014experimental}.  
Three measurement points (W1, W2, and W3) are assigned to record the free 
surface height, and three probes (P1, P2, and P3) are employed to capture the 
pressure signals.
We consider an inviscid flow with a density of $\rm \rho_{0}=1000 kg/m^{3}$ and
a gravitational constant of $\rm g=9.8 m/s^{2}$. 
According to the shallow water theory \cite{ritter1892fortpflanzung}, the 
maximum velocity is estimated as $2\sqrt{gH}$ to determine the speed of sound, 
where $H$ represents the initial water depth.

\begin{figure}[htb!]
	\centering
	\includegraphics[width=\textwidth]{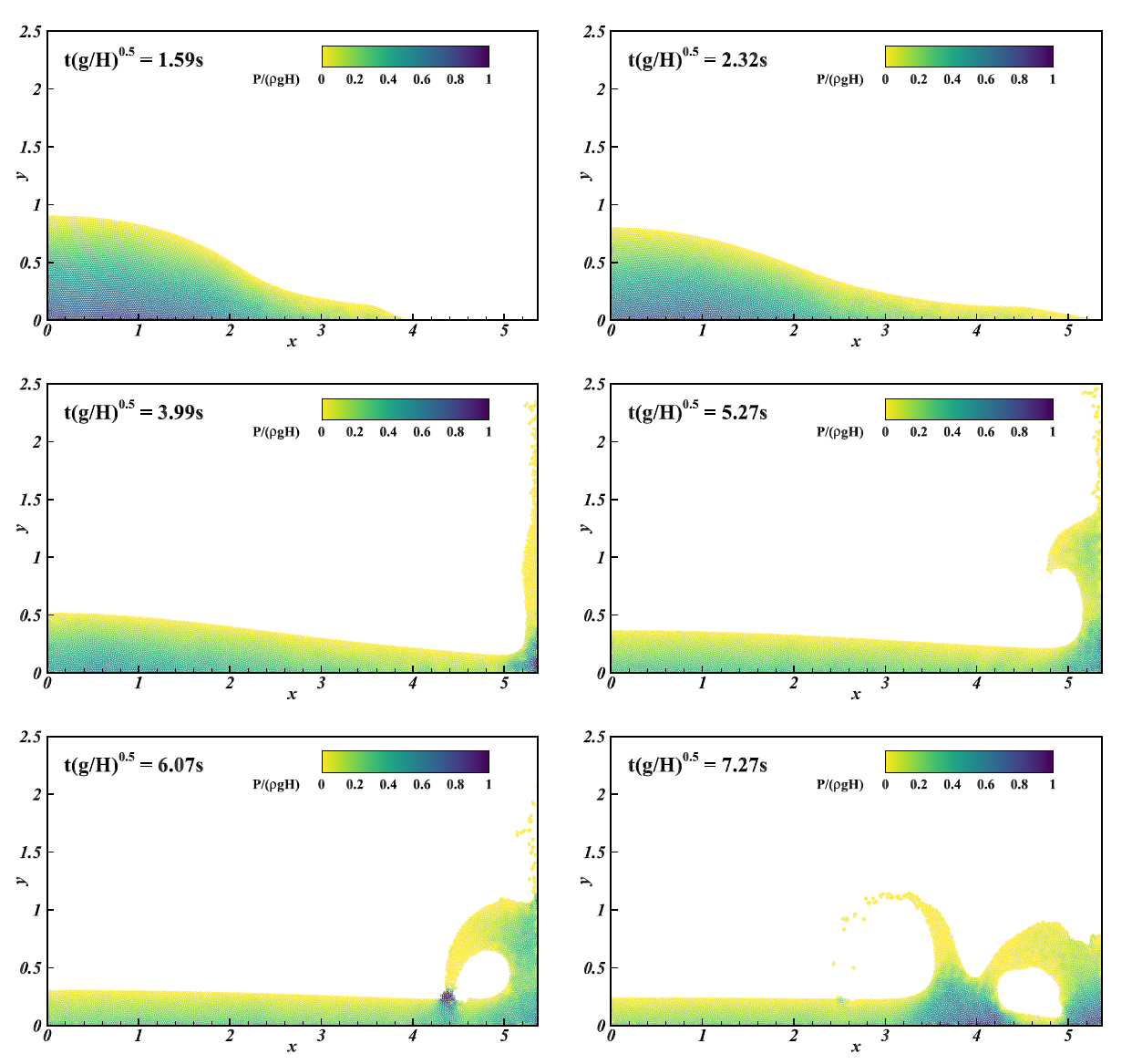}
	\caption{Dam-break flow: Snapshots of particles and pressure 
		distributions during the time evolution with $H/\Delta x = 60$.}
	\label{dambreakfreesurface}	
\end{figure}

\begin{figure}[htb!]
	\centering
	\includegraphics[width=\textwidth]{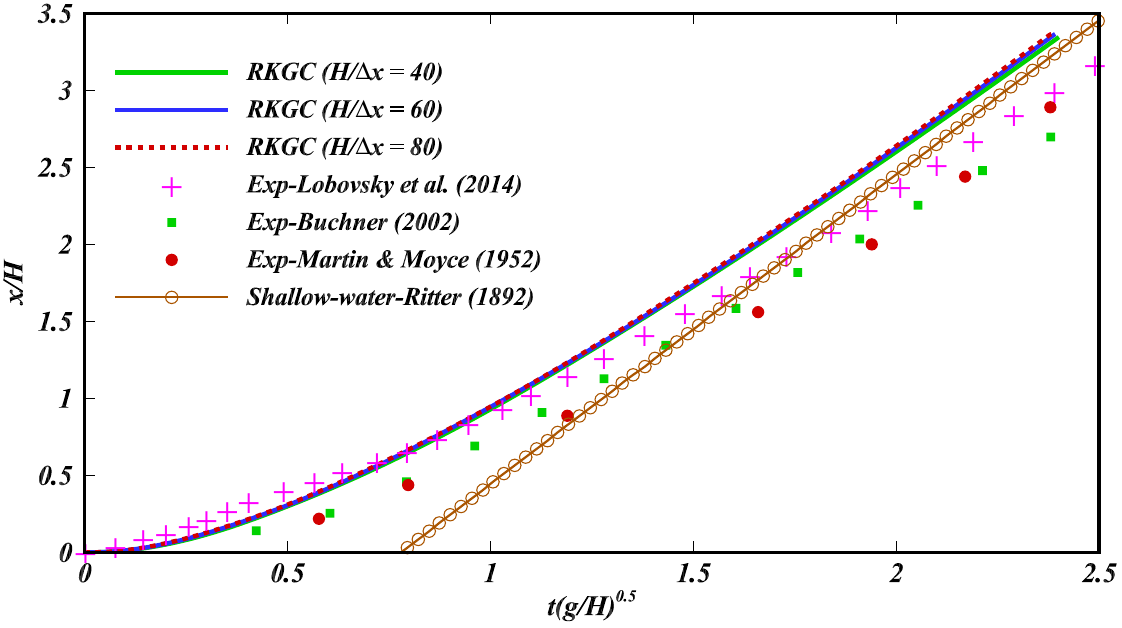}
	\caption{Dam-break flow: Time evolution of the surge-wave front.}
	\label{dambreak2DWaveFront}	
\end{figure}

Fig. \ref{dambreakfreesurface} shows several typical snapshots of the time 
evolution of the free surface obtained by the RKGC formulation.
The obtained results demonstrate smooth pressure distributions and robust 
free-surface profiles and align well with experimental observations 
\cite{lobovsky2014experimental} and previously reported simulation 
results \cite{zhang2017weakly, zhang2020dual, ren2023efficient}.
RKGC could appropriately capture key flow characteristics, including high 
roll-up along the downstream wall, a prominently reflected jet, and free 
surface disruption caused by the arrival of the secondary wave.
Fig. \ref{dambreak2DWaveFront} shows the predicted propagation of the 
surge-wave front, along with comparisons to data measured in various 
experiments \cite{buchner2002green, martin1952experimental, 
	lobovsky2014experimental} and the analytical solution derived from 
the shallow-water equation \cite{ritter1892fortpflanzung}.
It is observed that the present predictions have good convergence and agree 
well with the experiments before $t\sqrt{gH} < 1$, aligning closely with the 
analytical solution afterward, but overestimate the front speed obtained from 
the experiments, which was also observed in other simulations \cite{ferrari2009new, 
	adami2012generalized, zhang2017weakly}.

\begin{figure}[htbp]
	\centering
	\vspace{-2.2cm}
	\includegraphics[width=\textwidth]{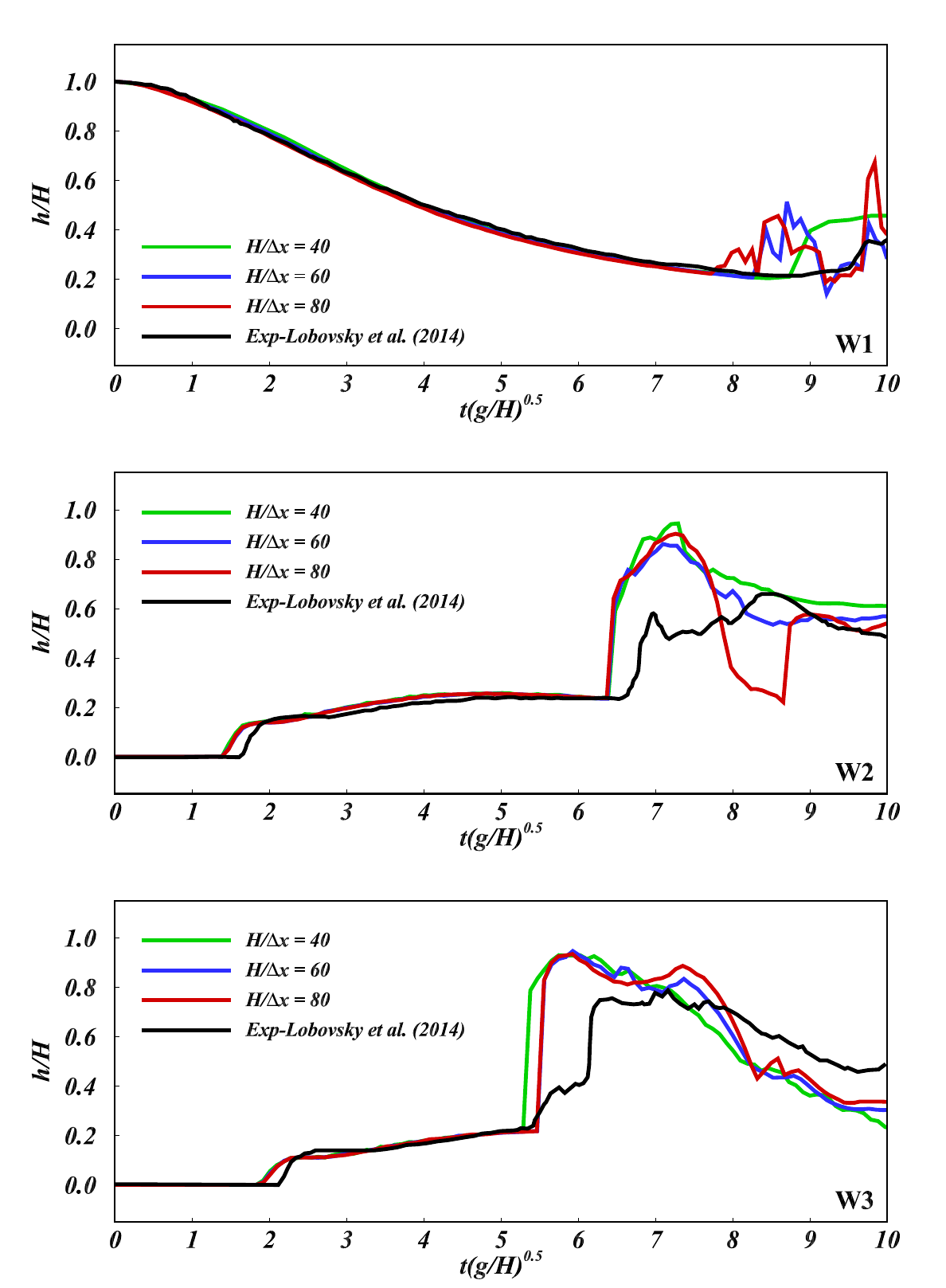}
	\caption{Three-dimensional dam-break: water levels recorded 
		at W1, W2, and W3. Convergence study and comparison against 
		experimental results presented by Lobovsky et al. 
		\cite{lobovsky2014experimental}.}
	\label{dambreakwaveheight}
\end{figure}

\begin{figure}[htb!]
	\centering
	\vspace{-2.8cm}
	\includegraphics[width=\textwidth]{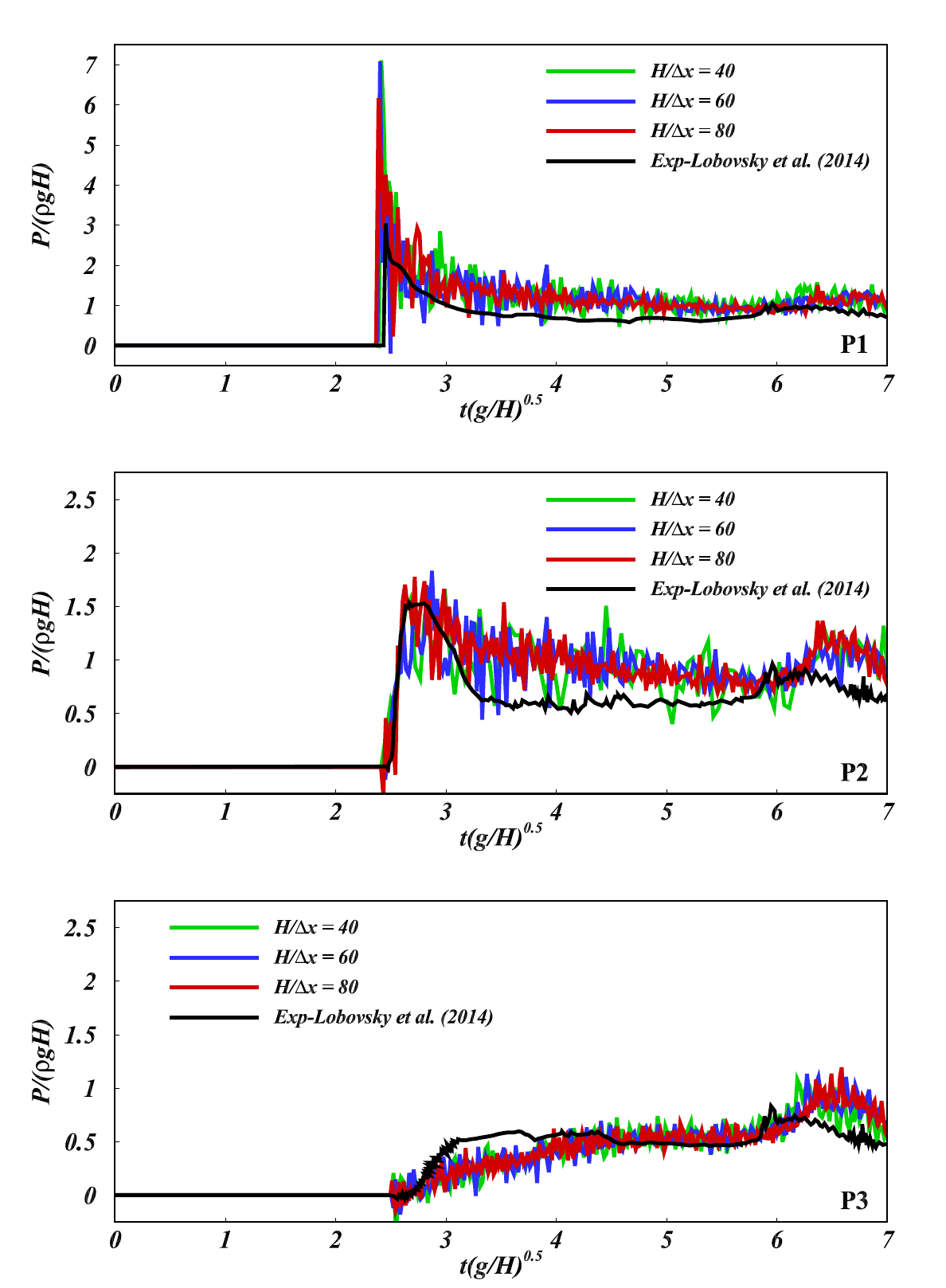}
	\caption{Three-dimensional dam-break: history of pressure 
		signals recorded at probes P1, P2 and P3. Convergence 
		study and comparison against experimental results
		presented by Lobovsky et al. \cite{lobovsky2014experimental}.}
	\label{dambreakpressure}
\end{figure}

The comparisons of the water levels recorded at W1, W2, and W3 with experimental 
observations obtained from Ref. \cite{lobovsky2014experimental} are presented in 
Fig. \ref{dambreakwaveheight}.
The wave height exhibits good agreement with the experimental data. 
Nonetheless, we observe discrepancies in higher run-up waves (W1) and 
marginally faster wavefront (W2 and W3) in the current results.
Similar observations were also reported in previous numerical studies 
\cite{ferrari2009new, zhang2017weakly, ren2023efficient}, potentially 
attributed to the adoption of inviscid flow in the current study, leading 
to violent wave breaking up and splashing.
The history of pressure signals recorded at P1, P2, and P3 is presented 
in Fig. \ref{dambreakpressure}.
The current results have good agreement with the experimental observations 
\cite{lobovsky2014experimental}, except for observed pressure fluctuations 
in the current study resulting from the weakly compressible assumption, 
which tend to decrease with increasing spatial resolutions. 
Discrepancies in pressure magnitudes at P2 and P3 were also reported 
in other studies \cite{cercos2015aquagpusph, zhang2020dual} where different 
WCSPH methods were employed.
The overestimated pressure peak at P2 by the current method is potentially 
due to the weakly compressible model. 
In addition, the air cushion effect in the experiment may also decrease its 
pressure peak \cite{lobovsky2014experimental}.
The occurrence time of the pressure observation at P3 aligns with 
$t\sqrt{g/H}\approx 2.7$ very well, though the peak value is underestimated.

\begin{figure}[htbp]
	\centering
	\subfigure[]{\includegraphics[width=\textwidth]
		{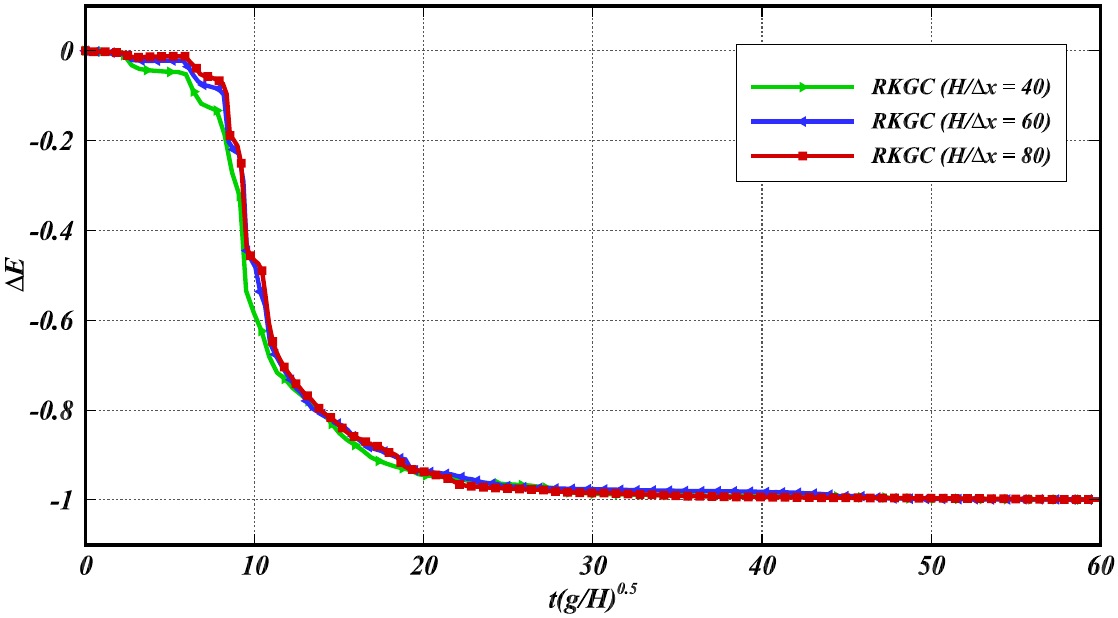}
		\label{dambreak2DTKERKGC}}
	\subfigure[]{\includegraphics[width=\textwidth]
		{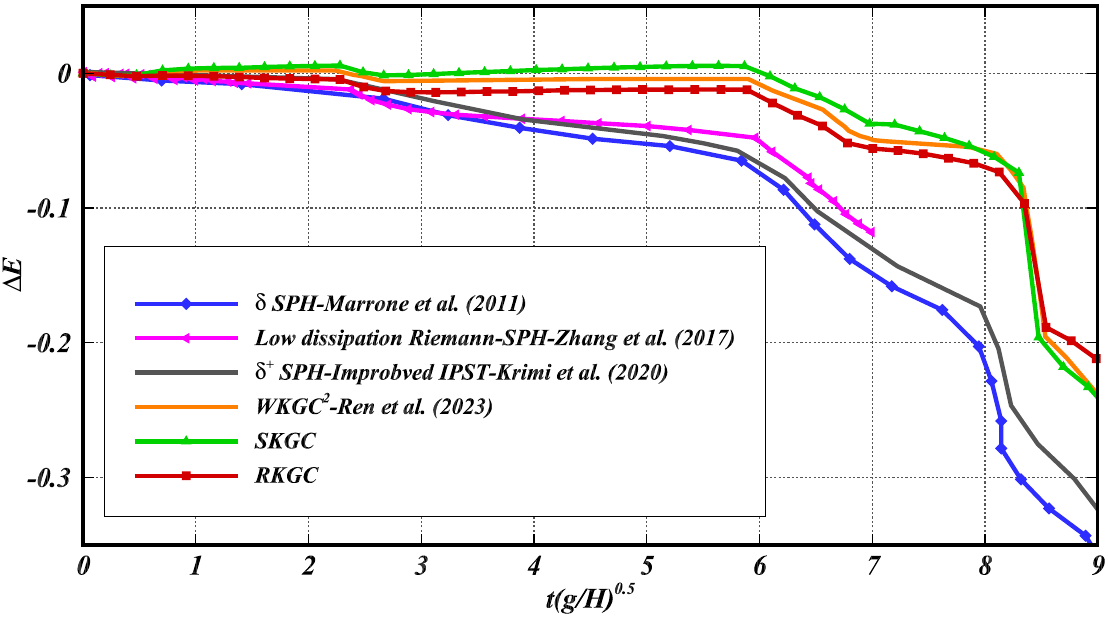}
		\label{dambreak2DTKE}}
	\caption{Dam-break flow: evolution of mechanical energy. 
		(a) Global evolution; 
		(b) Zoom in on the initial evolution and compare with 
		references ($H/\Delta x = 80$). }
	\label{dambreak2dtotalkineticenergydecay}
\end{figure}

Following the definition of the dissipation of total mechanical energy 
\cite{marrone2011delta}:
\begin{equation}
	\Delta E = \frac{E-E_{0}}{E_0-E_{\infty}}
	\label{mechanicalenergydissipation},
\end{equation}
where $E$ the total mechanical energy, $E_{0}$ is the initial mechanical energy, 
and $E_{\infty}$ is the mechanical energy after reaching the equilibrium state.
Fig. \ref{dambreak2dtotalkineticenergydecay} displays the evolution of the 
mechanical energy of RKGC, and compares it with SKGC and different references.
Fig. \ref{dambreak2DTKERKGC} shows that with increased resolutions, the 
numerical dissipation decreases rapidly.
It is observed in Fig. \ref{dambreak2DTKE} that RKGC has lower energy decay 
compared to other numerical results \cite{marrone2011delta, zhang2017weakly, 
	krimi2020wcsph}, where no correction was employed.
SKGC results in energy increasing before the splashing ($t\sqrt{gH}<6$), and 
involving the extra weighting of the KGC employed by Ren et al. 
\cite{ren2023efficient} could alleviate this issue, but it still leads to an 
energy increase before the roll-up wave ($t\sqrt{gH}<2.3$).
However, RKGC could maintain the energy before the splashing and leads to a 
slightly lower energy decay rate afterwards compared to SKGC and the results 
in Ref. \cite{ren2023efficient}, except for the energy-increasing artifacts 
observed for the latter.
\subsection{Three-dimensional oscillating wave surge converter (OWSC)}
\begin{figure}[htbp]
	\centering
	\includegraphics[width=\textwidth]{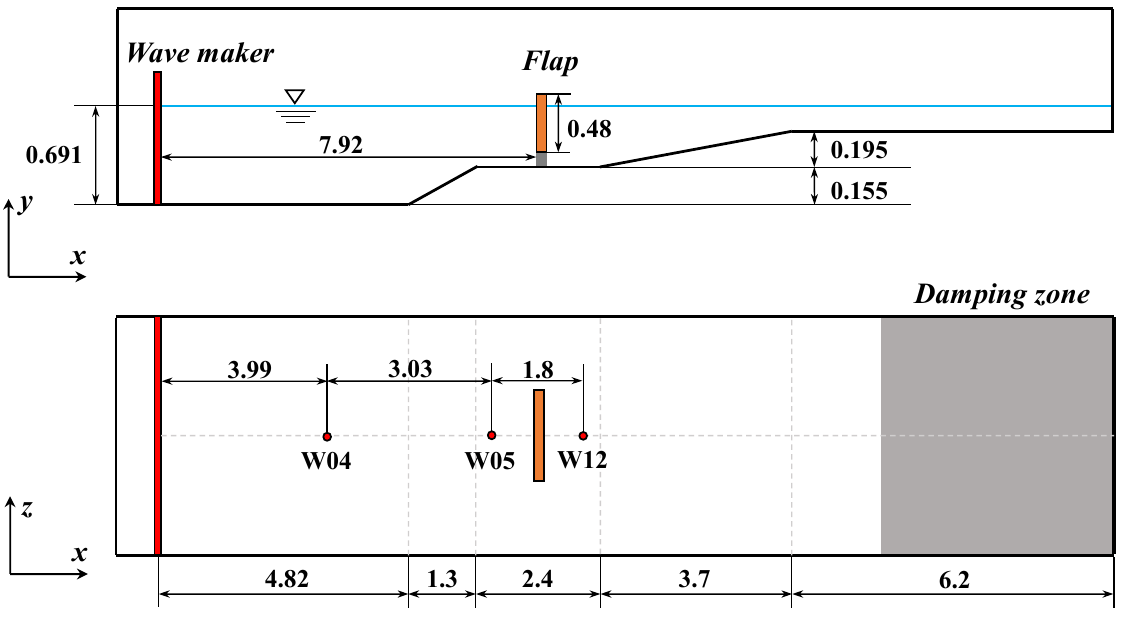}
	\caption{OWSC: Schematic depiction of the wave tank and the OWSC model.}
	\label{owscconfiguration}	
\end{figure}

As a forefront wave energy converter, the oscillating wave surge converter 
(OWSC) has showcased remarkable energy absorption capabilities and hydrodynamic 
performance, and it has been widely studied with numerical and experimental methods 
\cite{rafiee2013numerical, wei2015wave, brito2016coupling, zhang2021efficient}.
In this section, the three-dimensional OWSC is investigated with the RKGC formulation.
Fig. \ref{owscconfiguration} illustrates the configurations of the wave tank 
and the OWSC model, which are identical to the experimental setup detailed in 
Ref. \cite{wei2015wave}. 
The wave tank measures 18.42 m in length, 4.58 m in width, and 1.0 m in height. 
The OWSC device is simplified as a flap with dimensions of 0.48 m in height, 
1.04 m in width, and 0.12 m in thickness, and is hinged to a base with a height 
of 0.16 m. 
The flap has a mass of 33 kg, and its angular inertia is 1.84 $\rm kg/m^{2}$.
To measure the wave elevation and impact pressure on the flap, three wave gauges, 
as shown in Fig. \ref{owscconfiguration}, and six pressure sensors, whose positions 
are listed in Table \ref{owscpressureposition}, are employed.
Note that the present model is 1:25 scaled to that in the Ref. \cite{wei2015wave}, 
and all the results present in this work have been converted to the full scale 
accordingly.
\begin{table}[htb!]
	\small
	\renewcommand\arraystretch{1.0}
	\centering
	\captionsetup{font={small}}
	\caption{OWSC: Pressure sensor positions on the front flap along the 
		z-axis from the device's center, with $y=0$ representing the mean 
		water level.}
	\begin{tabularx}{14cm}{@{\extracolsep{\fill}}cccccc}
		\hline
		\quad No. & y-axis(m) & z-axis(m) & No. & y-axis(m) 
		& z-axis(m) \quad \\
		\midrule
		\quad P01 & -0.046 & 0.468 & P09 & -0.117 & 0.156 \quad \\
		\quad P03 & 0.050  & 0.364 & P11 & 0.025  & 0.052 \quad \\
		\quad P05 & -0.300 & 0.364 & P13 & -0.239 & 0.052 \quad \\
		\bottomrule
	\end{tabularx}
	\label{owscpressureposition}
\end{table}

Considering the regular wave with a height $H=5$m and a period $T=10$s
at full scale, a piston-type wave maker is employed to generate regular 
waves, adopting an ensemble of dummy particles whose motion follows the 
linear wavemaker theory \cite{dean1991water}, where the particle 
displacement in $x$-direction $x_{a}$ is determined by 
\begin{equation}
	x_{a} = S \sin(ft+\phi),
	\label{wavemakertheory}
\end{equation}
with $S$ the wave stroke, $f$ the wave frequency, and $\phi$ the initial 
phase. The wave stroke is further defined as 
\begin{equation}
	S=\dfrac{H \sinh(2kh_{0})+2kh_{0}}{\sinh(2kh_{0})\tanh(kh_{0})},
	\label{wavestroke}
\end{equation}
where $h_{0}$ is the water depth, and $k$ is the wave number.
To minimize wave reflection effects, a damping zone \cite{lind2012incompressible}
as illustrated in Fig. \ref{owscconfiguration}, is established at the 
end of the wave tank.  
Within this damping zone, the particle velocity undergoes decay according to
\begin{equation}
	\boldsymbol{\rm v}=\boldsymbol{\rm v_{0}}
	\left(1.0-\Delta t \theta\left(\dfrac{\boldsymbol{\rm r
	-r_{0}}}{\boldsymbol{\rm r_{1} - r_{0}}}\right)\right).
	\label{owscdampingzonedecay}
\end{equation}
Here, $\boldsymbol{\rm v_{0}}$ the initial velocity of the fluid particle 
at the entry of the damping zone, $\boldsymbol{\rm v}$ the velocity after 
damping, $\Delta t$ the time step, and $\boldsymbol{\rm r_{0}}$ and 
$\boldsymbol{\rm r_{1}}$ are the initial and final positions of the damping 
zone, respectively.
The reduction coefficient $\theta=5.0$ governs the modifications on the 
velocity at each time step in the current simulation.
The entire system is discretized with a particle spacing of 0.03 m, 
resulting in 1.542 million fluid particles and 0.628 million solid particles.
The present numerical results have been compared with both experimental 
observations and numerical investigations reported in Ref. \cite{wei2015wave}.

\begin{figure}[htb!]
	\centering
	\makebox[\textwidth][c]{
		\subfigure[]{\includegraphics[width=.5\textwidth]
			{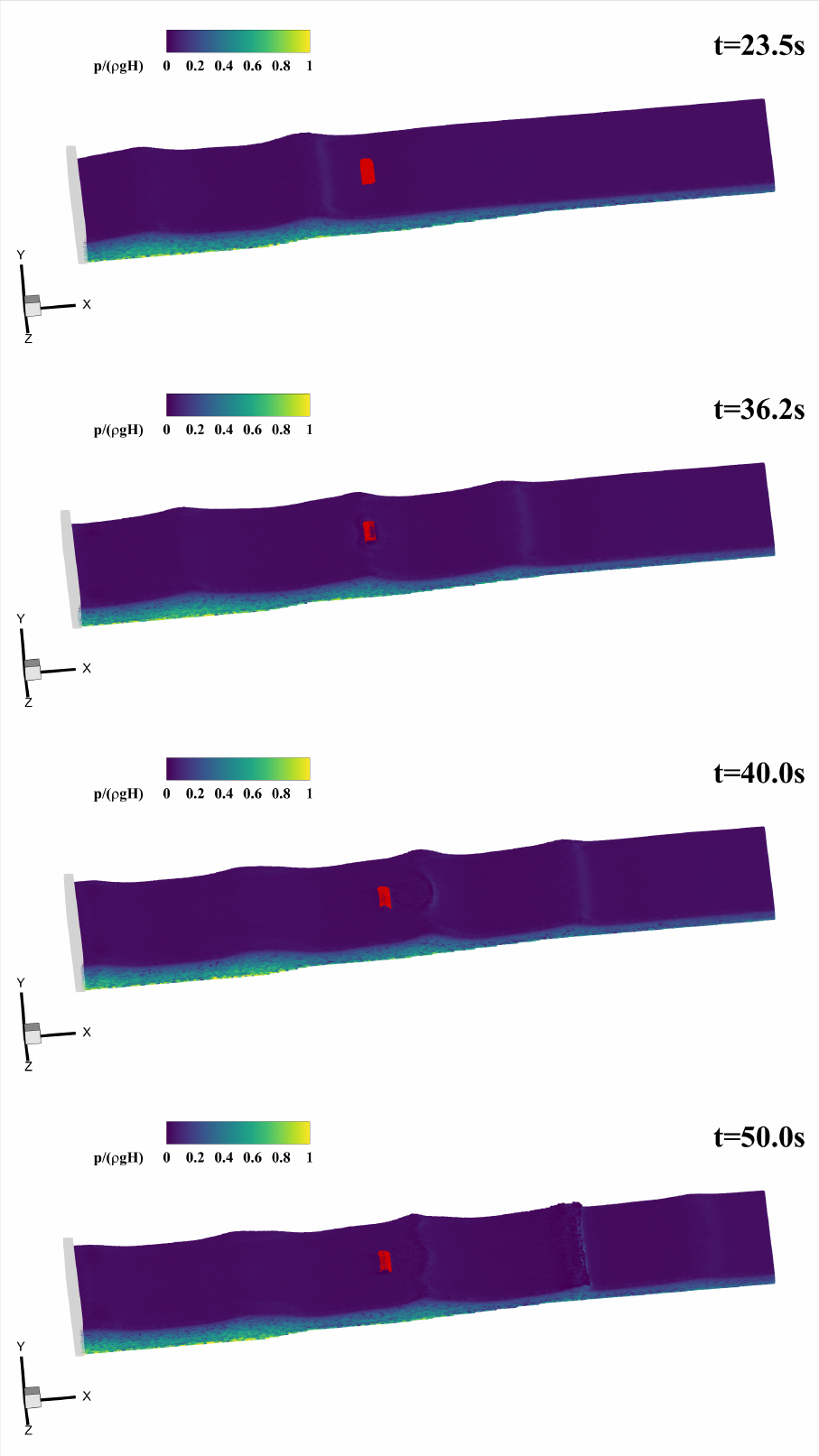}\label{owscsnapshotpressure}}
		\subfigure[]{\includegraphics[width=.5\textwidth]
			{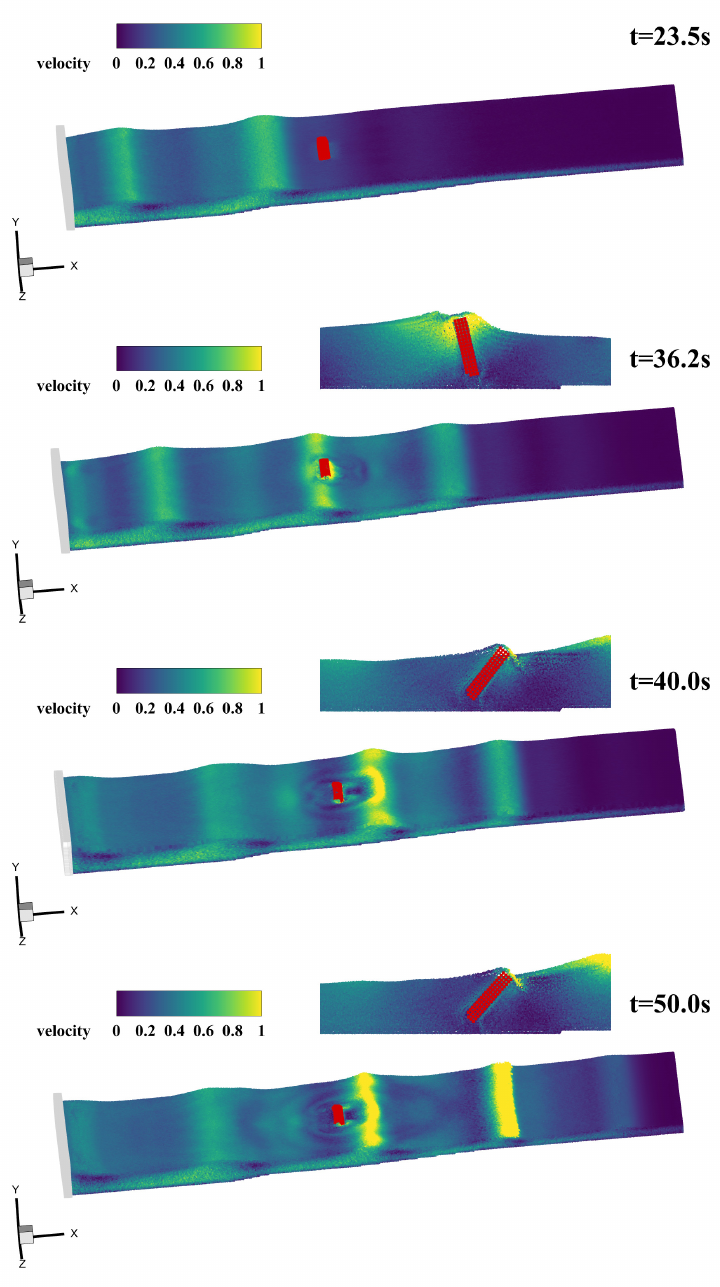}\label{owscsnapshotvelocity}}}
	\caption{OWSC: Snapshots of free surfaces and the flap 
		motion during time evolution. 
		(a) Fluid particles are colored by the normalized pressure; 
		(b) Fluid particles are colored by the velocity magnitude.}
	\label{owscsnapshot}
\end{figure}

Fig. \ref{owscsnapshot} displays snapshots of the free-surface profile colored 
by the normalized pressure \ref{owscsnapshotpressure} and velocity magnitude 
\ref{owscsnapshotvelocity}, respectively.
The results clearly demonstrate that the current method effectively captures the 
dynamic free-surface elevation, including wave reflection and breaking around the flap.
Additionally, the outcomes exhibit smooth pressure and velocity fields, even during 
the intensive wave interactions around the flap, where wave reflection and breaking 
are observed.
Furthermore, the cross-sectional slices along the middle line provide insights into 
the rotational state of the flap.
These observations are consistent with those reported in the references 
\cite{wei2015wave, zhang2021efficient}, as evidenced by the wave height and flap 
rotation angle history presented below.

\begin{figure}[htbp]
	\centering
	\vspace{-2cm}
	\includegraphics[width=\textwidth]{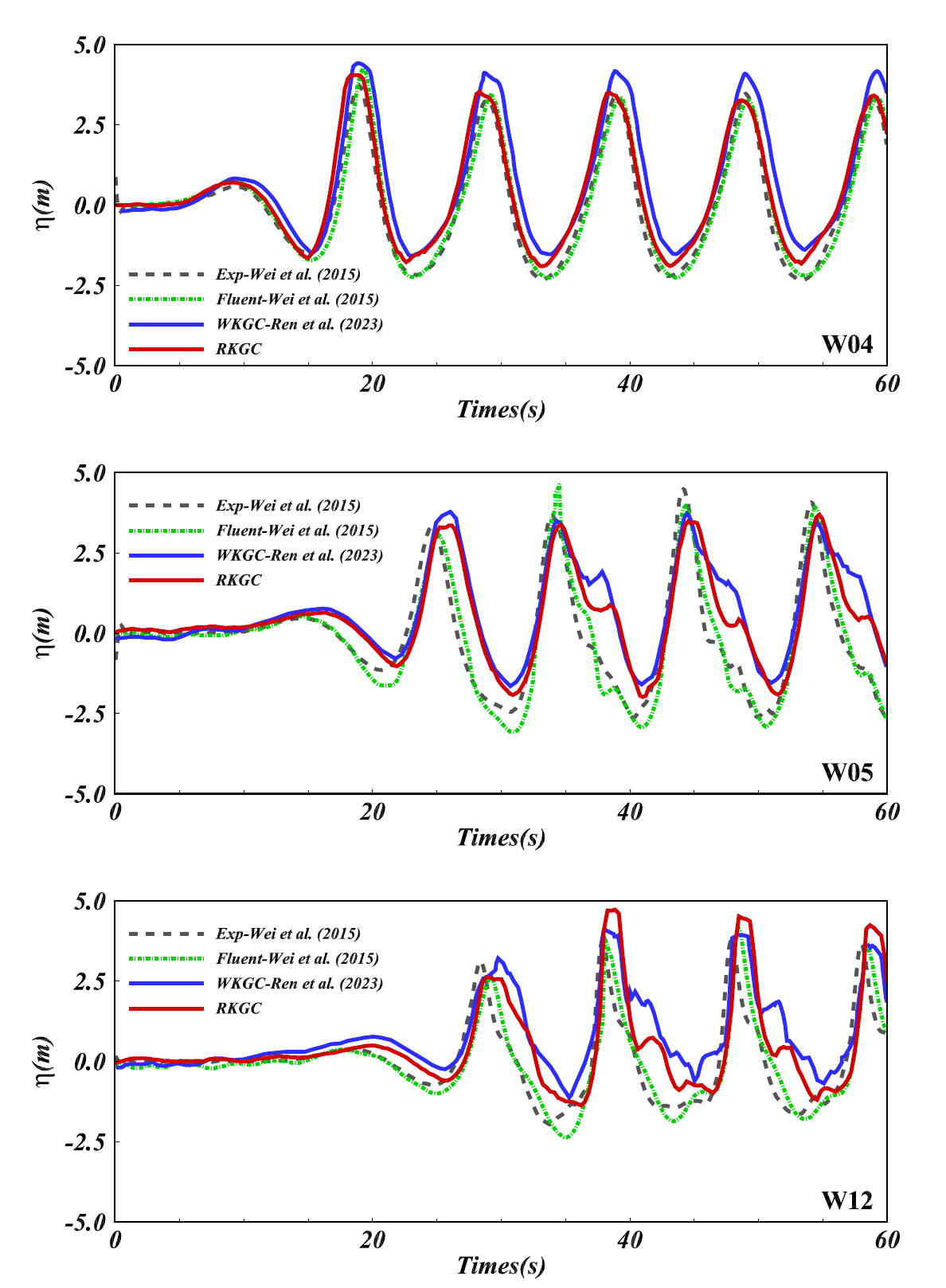}
	\caption{OWSC: Comparison of free surface elevations for wave height $H=5.0m$ 
		and period $T=10.0s$ against results of Wei et al. \cite{wei2015wave} and 
		Ren et al. \cite{ren2023efficient}.}
	\label{owscwaveheight}
\end{figure}

The observed wave evaluations are depicted in Fig. \ref{owscwaveheight}, offering 
a comparison with the reference results \cite{wei2015wave, ren2023efficient}.
The RKGC formulation demonstrates good agreement with the results obtained from 
experiments, particularly at locations W04 and W05, which are in the seaward direction 
from the flap. 
However, discrepancies, especially the overestimation of wave crest height, are 
noticeable at location W12, positioned behind the flap.
These differences may arise from wave reflection and breaking around the flap. 
Additionally, the absence of a turbulence model in the current study also contributes 
to these observations, as the reference numerical results \cite{wei2015wave} utilize 
a turbulence model, introducing additional numerical dissipation. 
Compared to the results in Ref. \cite{ren2023efficient} where WKGC$^2$ is adopted, 
RKGC shows better alignment with experimental, particularly predicting more accurate 
wave heights at W04 and W05 positions and wave falls at all three positions.

\begin{figure}[htb!]
	\centering
	\includegraphics[width=\textwidth]{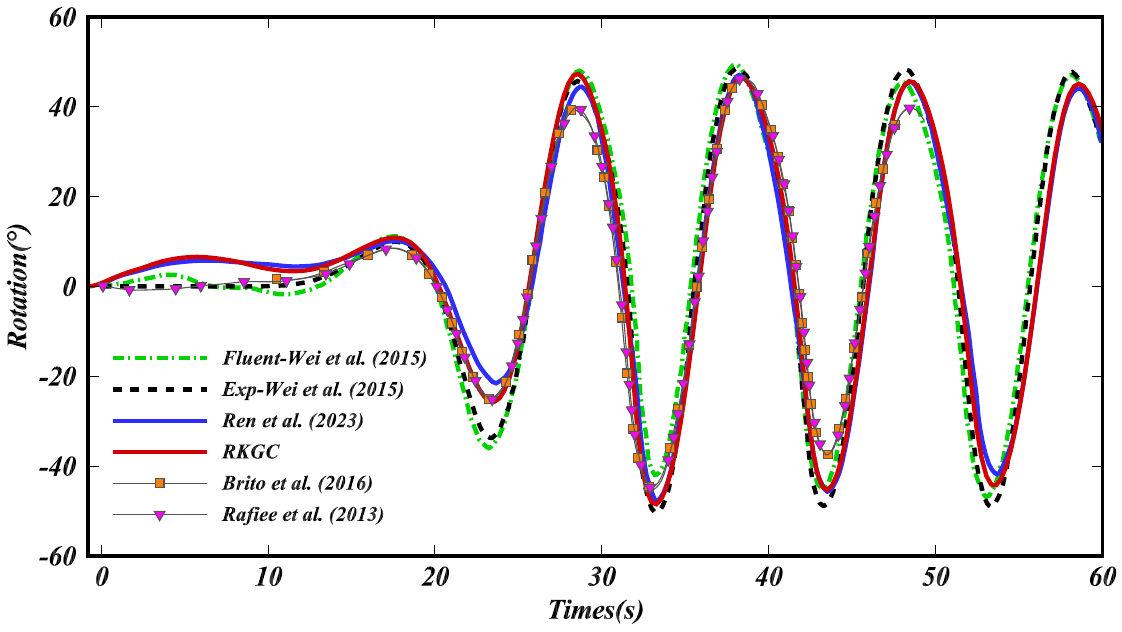}
	\caption{OWSC: Comparison of the time evolution of the flap rotation.}
	\label{owscrotationrate}
\end{figure}

Fig. \ref{owscrotationrate} presents the rotation angle of the flap, providing a 
comparison with the experimental observations and other numerical predictions. 
The comparison underscores the good agreement with the experimental and Fluent 
results \cite{wei2015wave}.
While predictions obtained by standard SPH formulations without corrections, such 
as those reported in Ref. \cite{rafiee2013numerical, brito2016coupling}, often 
underestimate the extreme rotation angle of the flap compared to experimental results,
the prediction gained with the RKGC formulation and WKGC$^2$ \cite{ren2023efficient} 
notably overcomes this limitation.
They provide more accurate predictions for flap rotating angles, aligning more closely 
with the experimental observations.
The RKGC formulation in this study is applied only in the fluid domain rather than the 
fluid-structure interaction, thus, it gives slightly closer predictions to experiments 
than that in Ref. \cite{ren2023efficient}. 

\begin{figure}[]
	\vspace{-2cm}
	\begin{subfigure}{}
		\includegraphics[width=\textwidth]{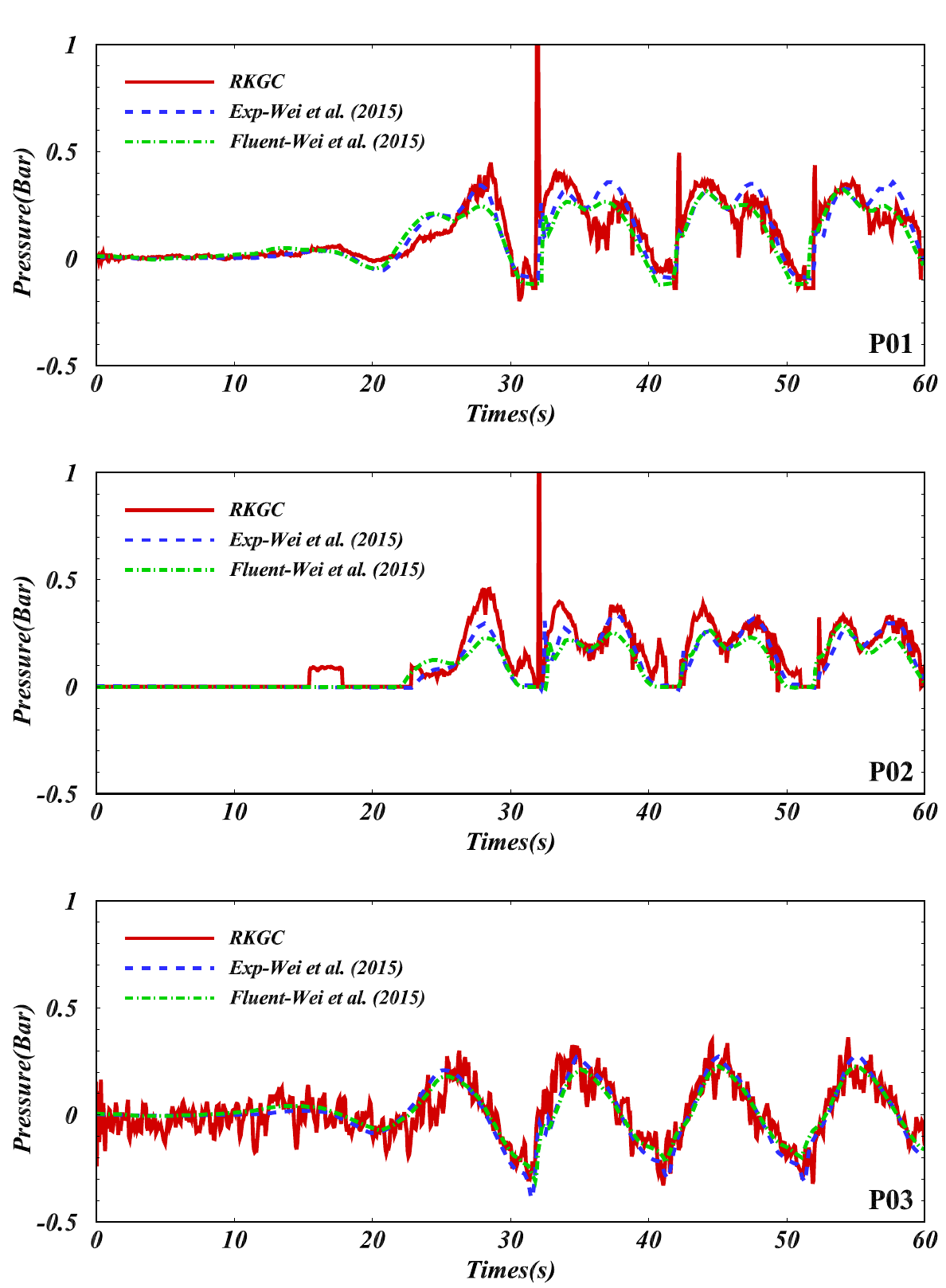}
	\end{subfigure}
    \caption{}
\end{figure}  
\begin{figure}[]
	\vspace{-2cm}
	\ContinuedFloat
	\begin{subfigure}{}
		\includegraphics[width=\textwidth]{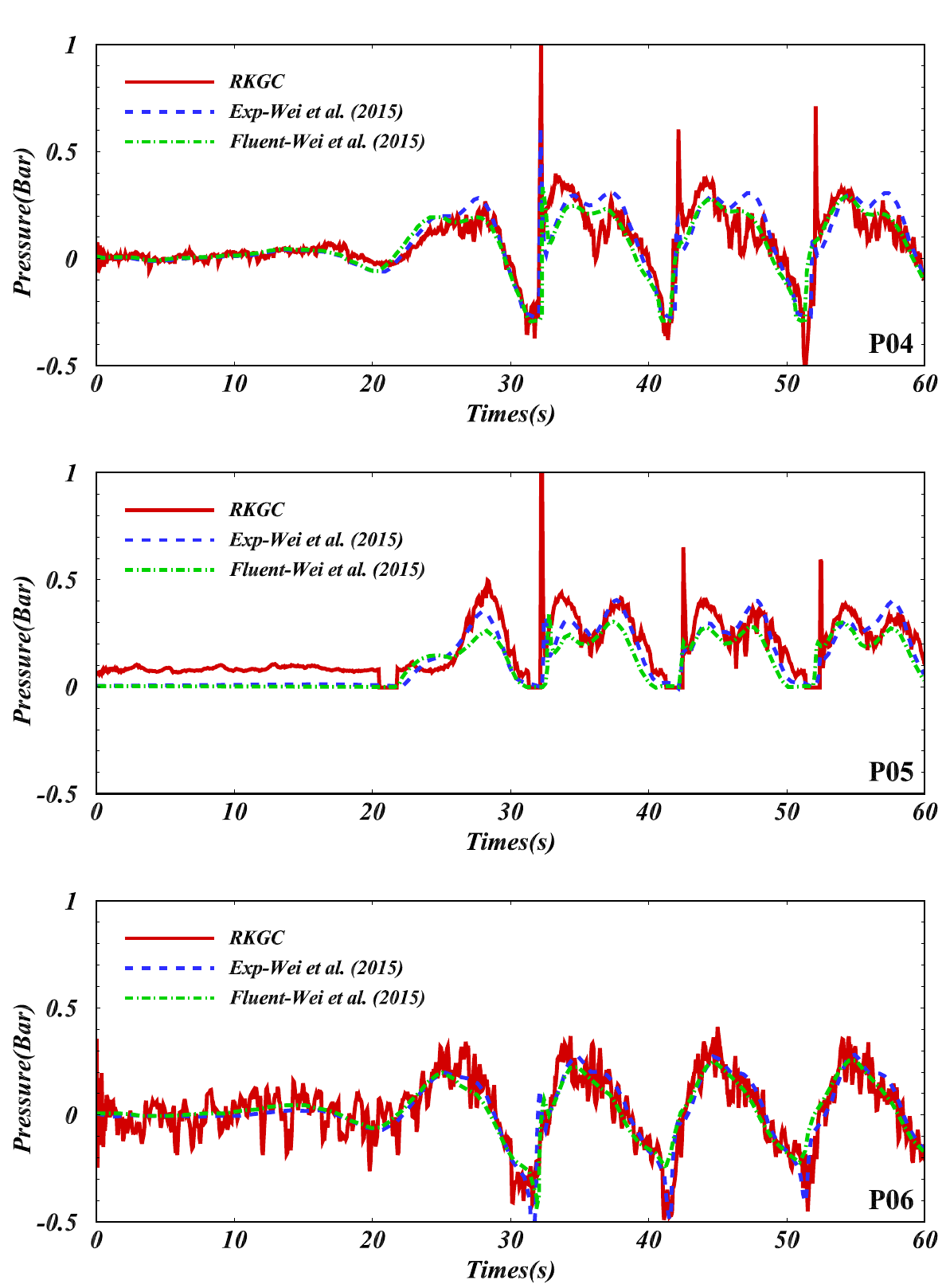}
	\end{subfigure}
	\caption{OWSC: Comparison of wave load time histories on the flap 
		for $H=5.0m$ and $T=10.0s$ against the results of Wei et al. 
		\cite{wei2015wave}.}
	\label{owscpressure}
\end{figure}
The pressure evolution over time at each probe on the flap is illustrated and 
compared in Fig. \ref{owscpressure}. 
The results obtained using the RKGC formulation align well with the experimental 
data, predicting reasonable slamming pressures despite some pressure oscillations 
due to the weakly compressible assumption. 
Compared to the pressure results in Ref. \cite{zhang2021efficient, ren2023efficient} 
based on the SPH method, the current observed pressure shows fluctuations due to the 
density reinitialization method employed in Ref. \cite{salis20243d}, which has been 
proven more suitable for free-surface problems.
Moreover, the RKGC method predicts pressures that are more consistent with experimental 
results than those obtained using Fluent. 
This includes capturing higher double pressure peaks at P01, P02, and P05, as well as 
the lower pressure drops at P03 and P06.
However, some discrepancies still exist, which may be attributed to air entrainment in 
the splash passing the flap, a factor not considered in the current method.
%
%
\section{Extension}\label{extension}
Indeed, the concept of the reverse KGC formulation can be readily 
extended to accommodate second- or even higher-order consistency 
conditions of the SPH gradient approximation.
This extension uses the correction function for RKPM proposed 
by Liu et al. \cite{liu1995reproducing, liu1995reproducing2}, 
provided that the corresponding particle relaxation is employed.
The SPH gradient approximation in the non-conservative form can 
be expressed similarly to Eq. \eqref{strong-corrected} as 
\begin{equation}
	\nabla \psi_{i}=\sum_j\psi_{ij}
	\boldsymbol{C_{i}}\left(\boldsymbol{\rm r}_{i},\boldsymbol
	{\rm r}_{j}\right)\nabla W_{ij}V_{j}
	\label{highorderreproduce},
\end{equation}
where $\boldsymbol{C_{i}}\left(\boldsymbol{\rm r}_{i},
\boldsymbol{\rm r}_{j}\right)={C}_{0}
\left(\boldsymbol{\rm r}_{i}\right)+\boldsymbol{C_{1}}
\left(\boldsymbol{\rm r}_{i}\right)\left(\boldsymbol{\rm r}_{j}-
\boldsymbol{\rm r}_{i}\right)$ represents the correction function (see 
Refs. \cite{liu1995reproducing, liu1995reproducing2} for more definitions).
Following Taylor expansion of $\psi_{j}$ and substituting it into 
\eqref{highorderreproduce}, one can obtain
\begin{equation}
	\nabla \psi_{i}=-\sum_{j}\left(\nabla \psi_{i}
	\cdot\boldsymbol{\rm r}_{ij}
	+\frac{1}{2}\nabla\cdot\nabla\psi_{i}:
	\boldsymbol{\rm r}_{ij}
	\otimes\boldsymbol{\rm r}_{ij}\right)
	\boldsymbol{C_{i}}\left(\boldsymbol{\rm r}_{i},
	\boldsymbol{\rm r}_{j}\right)
	\nabla W_{ij}V_{j}.
\end{equation}
To vanish the leading moments and ensure second-order consistency, 
the following conditions should be satisfied simultaneously:
\begin{equation}
	\begin{cases}
		\displaystyle\sum_{j}-\boldsymbol{\rm r}_{ij}\otimes
		\boldsymbol{C_{i}}\left(\boldsymbol{\rm r}_{i},
		\boldsymbol{\rm r}_{j}\right)
		\nabla W_{ij}V_{j}=\mathbf{I}\\
		\displaystyle\sum_{j}\boldsymbol{\rm r}_{ij}
		\otimes\boldsymbol{\rm r}_{ij}\cdot
		\boldsymbol{C_{i}}\left(\boldsymbol{\rm r}_{i},
		\boldsymbol{\rm r}_{j}\right)\nabla W_{ij}V_{j}=\mathbf{0}
	\end{cases}.
	\label{correctiontensor}
\end{equation}
Subsequently, $C_{0}$ and $\boldsymbol{C_{1}}$ 
can be obtained by solving Eq. \eqref{correctiontensor}.
Consequently, the reverse-corrected conservative formulation 
with second-order consistency can be written similarly to Eq. 
\eqref{rkgc} as
\begin{equation}
	\nabla \psi_{i}=-\sum_{j}
	\left(\psi_{i}\boldsymbol{C_{j}}+\psi_{j}\boldsymbol
	{C_{i}}\right)\nabla W_{ij}V_{j}
	\label{highorderrkgc},
\end{equation}
which can be further rewritten as
\begin{equation}
	\nabla\psi_{i}=-\psi_{i}\sum_{j}\left(\boldsymbol{C_{i}}+
	\boldsymbol{C_{j}}\right)\nabla W_{ij}V_{j}+\sum_{j}\psi_{ij}
	\boldsymbol{C_{i}}\nabla W_{ij}V_{j}
	\label{highorderrkgcexplain}.
\end{equation}
The first term on the RHS can also be nullified by particle relaxation 
based on the correction function, while the second term accurately 
reproduces the gradient as indicated in Eq. \eqref{highorderreproduce}.
Therefore, the reverse-corrected conservative formulation in Eq. 
\eqref{highorderrkgc} could continuously exhibit second-order consistency.
However, obtaining the corresponding correction function and achieving 
convergence in the particle relaxation driven by the correction function 
to fulfill second-order consistency still presents significant challenges.
%
%
\section{Conclusion}\label{conclusion}
This paper introduces the reverse KGC (RKGC) formulation, which is 
conservative and, integrating the particle relaxation based on the KGC 
matrix, ensures the zero- and first-order consistencies without 
explicit dependence on smoothing length.
The implementations in typical SPH methods, including Lagrangian SPH 
and Eulerian SPH, exhibited considerably improved accuracy, especially 
good energy conservation properties in general free-surface problems. 
While acknowledging the ease of extending the scheme to high-order 
consistency, one expects future work to address the persistent challenges 
in achieving converged solutions for particle relaxation driven by the KGC 
matrix and high-order correction function, especially for three-dimensional 
complex geometries and the situation employing a $h/\Delta x$ value 
smaller than 1.0.
Additionally, extending RKGC for SPH solid dynamics and to a similar 
idea for Laplacian operators is to be considered as future work. 
%
%
%
\section*{Acknowledgments}
\addcontentsline{toc}{section}{Acknowledgement}
Bo Zhang acknowledges the financial support provided by the China 
Scholarship Council (No. 202006230071). 
X.Y. Hu expresses gratitude to the Deutsche Forschungsgemeinschaft 
(DFG) for sponsoring this research under grant number DFG HU1527/12-4. 
The corresponding code for this work is available on GitHub at 
\url{https://github.com/Xiangyu-Hu/SPHinXsys}.
\bibliographystyle{elsarticle-num}
\bibliography{consistency}
\end{document}